\documentclass[twocolumn,aps,prl,preprintnumbers,showpacs,nofootinbib]{revtex4}
\usepackage[utf8]{inputenc}
\usepackage[english]{babel}
\usepackage{amsmath}
\usepackage{amsfonts}
\usepackage{amssymb}
\usepackage{epsfig}
\usepackage{graphics,psfrag,rotating}
\usepackage{graphicx}
\usepackage{dcolumn}
\usepackage{bm}
\usepackage{epstopdf}
\usepackage{color}
\usepackage[usenames,dvipsnames,svgnames]{xcolor}
\usepackage[colorlinks=true,
            linkcolor=red,
            urlcolor=gray,
            citecolor=blue]{hyperref}
  \usepackage{hyperref}

\newcommand{\lsim}   {\mathrel{\mathop{\kern 0pt \rlap
{\raise.2ex\hbox{$<$}}}
 \lower.9ex\hbox{\kern-.190em $\sim$}}}
\newcommand{\gsim}   {\mathrel{\mathop{\kern 0pt \rlap
{\raise.2ex\hbox{$>$}}}
\lower.9ex\hbox{\kern-.190em $\sim$}}}

\def\3nab{\tilde{\nabla}}

\def\hsp5{\hspace{5mm}}

\def\case#1/#2{\textstyle\frac{#1}{#2}}

\def\ber {\begin{eqnarray}}
\def\eer {\end{eqnarray}}
\def\bea {\begin{eqnarray}}
\def\eea {\end{eqnarray}}

\def\bc {\begin{center}}
\def\ec {\end{center}}
\def\case#1/#2{\frac{#1}{#2}}

\newcommand{\bw}{\begin{widetext}}
\newcommand{\ew}{\end{widetext}}

\newcommand{\be}{\begin{equation}}
\newcommand{\bse}{\begin{subequation}}
\newcommand{\ese}{\end{subequation}}
\newcommand{\ee}{\end{equation}}
\newcommand{\eei}{\end{eqnarray}\indent\indent}
\newcommand{\ba}{\begin{array}}
\newcommand{\ea}{\end{array}}
\newcommand{\bal}{\begin{eqnarray}}
\newcommand{\eal}{\end{eqnarray}}

\def\case#1/#2{\textstyle\frac{#1}{#2} }

\newcommand{\bver}{\begin{verbatim}}
\newcommand{\ever}{\end{verbatim}}

\begin{document}


\title{Complete density perturbations in the Jordan-Fierz-Brans-Dicke theory}
\author{
J.~A.~R.~Cembranos$\,^{(a)}$\footnote{Email: cembra@fis.ucm.es},
A.~de la Cruz Dombriz$\,^{(b,c)}$\footnote{Email: alvaro.delacruzdombriz@uct.ac.za},
L.~Olano Garc\'ia$\,^{(a)}$\footnote{Email: leandroolano@estumail.ucm.es}
}
\affiliation{$^{(a)}$ Departamento de F\'{\i}sica Te\'orica I, Universidad Complutense de Madrid, E-28040 Madrid, Spain.}
\affiliation{$^{(b)}$ Instituto de Ciencias del Espacio (ICE/CSIC) and Institut d'Estudis Espacials de Catalunya (IEEC), Campus UAB, Facultat de Ci\`{e}ncies, Torre C5-Par-2a, 08193 Bellaterra (Barcelona) Spain}
\affiliation{$^{(c)}$ Astrophysics, Cosmology and Gravity Centre (ACGC) and
Department of Mathematics and Applied Mathematics, University of Cape Town, Rondebosch 7701, Cape Town, South Africa.}

\date{\today}

\begin{abstract}
In the context of scalar-tensor theories we study the evolution of the density contrast for Jordan-Fierz-Brans-Dicke theories in a Friedmann-Lema\^itre-Robertson-Walker Universe. Calculations are performed in the Einstein Frame with the cosmological background described as $\Lambda$-Cold Dark Matter ($\Lambda$CDM) and supplemented by a Jordan-Fierz-Brans-Dicke field. By using a completely general procedure valid for all scalar-tensor theories, we obtain the exact fourth-order differential equation for the density contrast evolution in modes of arbitrary size. In the case of sub-Hubble modes, the expression reduces to a simpler but still fourth-order equation that is then compared with the standard (quasistatic) approximation.  Differences with respect to the evolution as predicted by the standard Concordance $\Lambda$CDM model
are observed depending on the value of the coupling.
\end{abstract}


%
%
                             

\maketitle

\section{Introduction}

The present accelerated phase that the Universe seems to be experiencing nowadays 
\cite{Riess} constitutes an open and major problem in modern cosmology.
With the assumption of General Relativity (GR) as the gravitational underlying theory, standard Einstein
equations (EE) in either matter or radiation dominated Universe give rise to decelerated periods of expansion. These fluids
are unable by themselves  to violate the strong energy condition that would provide cosmological acceleration once that for sufficiently large scales
a homogeneous and isotropic Universe is assumed.
Thus some kind of mechanism to guarantee acceleration is required. These mechanisms are usually classified in two different ways. The first one considers that the total stress-energy tensor appearing on the right-hand side of the EE should be dominated at late times by a hypothetical negative pressure fluid usually dubbed dark energy (see \cite{Copeland} and references therein). The second approach consists of modifying the left-hand side of EE, thus modifying gravity itself and interpreting the acceleration as a geometrical effect rather than as a consequence of the inclusion of exotic fluids.

However, both points of view are mathematically equivalent since geometrical modifications can be interpreted as curvature fluids and therefore interpreted as dark energy contributions. Some examples of these mechanisms are provided by Lovelock theories \cite{Lovelock}, Gauss-bonnet models \cite{GB}, scalar-tensor theories \cite{ST} or vector-tensor theories \cite{VT}, gravitational theories derived from extra dimensional models \cite{XD}; supergravity models \cite{sugra}, disformal theories \cite{disformal} or Lorentz violating and CPT breaking models of gravity \cite{LV}.
With the exception of a cosmological constant $\Lambda$, scalar fields models of quintessence  \cite{CaldwellDaveSteinhardt} are the easiest way to add a new fluid (or field) in an attempt to explain the late-time cosmological acceleration. However, these models present several drawbacks such as potential violations of the equivalence principle \cite{Damour}.

Another paradigmatic example of geometrical modifications of the gravitational interaction are
the scalar-tensor theories which add a scalar field  to the gravitational interaction. This scalar field can be interpreted as a new kind of fluid which does not have to verify the usual energy conditions. In fact, the so-called $f(R)$ theories \cite{fR,Alvaro's-thesis}, where the usual Einstein-Hilbert gravitational action is replaced by a more general $f(R)$ term, can be understood as a kind of scalar-tensor theory.
One of the earliest works for developing an alternative to GR was conducted by Brans and Dicke and was related with some previous work of Jordan and Fierz \cite{JFBD}. This theory is usually referred to as Jordan-Fierz-Brans-Dicke theory (JFBD), or also Brans-Dicke theory (BD) (see \cite{Brans:2005ra} for a
recent review). The action for such theories can be written as
\begin{eqnarray}
S&=&\int\frac{{\rm d}^4x}{16\pi G_*}\sqrt{-g}\left[\varphi R-\frac{\omega_0}{\varphi}\partial^\nu\varphi\partial_\nu\varphi\right]
+S_M[g_{\mu\nu};\psi],\nonumber\\
&&
\label{actionJF}
\end{eqnarray}
where $G_*$ holds for the bare gravitational coupling constant, $R$ the scalar curvature associated to the metric $g_{\mu\nu}$, $\sqrt{-g}$ the determinant of the metric, $S_M$ the action corresponding to matter fields $\psi$ and the metric, $\varphi$ the scalar field and $\omega_0$ the coupling between the scalar field and the metric.
The main difference in \eqref{actionJF} with respect to GR is that the gravitational constant is in fact non-constant but dependent on the scalar field $\varphi$. This scalar field contributes to the Lagrangian density with its own kinetic term. In addition, it can be shown that the evolution of the scalar field has as a source term the contracted matter stress-energy tensor. Therefore, making the scalar field dependent on the mass distribution and subsequently making the gravitational {\it constant} dependent of the mass as well. This effect provides a manifestation of the Mach's principle as interpreted by Dicke \cite{Brans:2005ra}.

JFBD theories are included in a more general set of the so-called scalar-tensor theories. These theories allow for a constant $\omega_0$ depending on the scalar field itself and may also include a potential to the scalar field Lagrangian density. Scalar-tensor theories are usually formulated in two different frameworks, the Jordan Frame (JF) and the Einstein Frame (EF). The former defines lengths and times as would be measured by ordinary laboratory apparatus so that all observables (time, redshift, among others) have their standard interpretation in this frame. The metric is minimally coupled to matter in the JF and the scalar field is coupled with the Ricci curvature. Action \eqref{actionJF} is defined in this frame. However, it is easier to work in the EF. This frame has the advantage of diagonalizing the kinetic terms for the spin-0 (the scalar field) and spin-2 (the graviton) degrees of freedom so that the mathematical consistency of the solutions of the theory are more easily discussed. In this case, the scalar field is coupled with matter \cite{Olive,Gilles}.

There are also strong theoretical arguments to take into account scalar-tensor theories. Among others, these arguments include the fact  that scalar partners of the graviton naturally arise in most attempts to quantise gravity or unify it with other interactions. Also, as we previously mentioned, other theories of gravity, such as $f(R)$ theories, can be expressed as scalar-tensor theories \cite{Alvaro's-thesis}. A different line of reasoning claims that the coupling between the scalar field and the matter density could provide a mechanism to alleviate the coincidence problem \cite{Chimento:2003iea}.

Once the late-time acceleration has been generated, the most important question to address is how to discriminate between competing dark energy models or theories that mimic the cosmological evolution as predicted by the $\Lambda$CDM or {\it Concordance Model} \cite{LCDM}.
Studies of the cosmic expansion history through high-redshift Hubble diagrams from SNIa \cite{SNIa}, baryon acoustic oscillations \cite{BAO} or CMB shift factor \cite{CMB_shift} cannot settle the underlying nature of dark energy by themselves, due to the fact that different theories can provide the same global expansion properties in a Friedmann-Lema\^itre-Robertson-Walker (FLRW) Universe  \cite{Linder:2005in}. Therefore, other types of cosmological and astrophysical observations are required in order to break this degeneracy \cite{degeneracy_references}. At this point it is necessary to study the growth of structures and extend the standard theory of cosmological perturbations from GR to more general gravity theories since these calculations can provide additional information and help to distinguish between similar expansion histories.

Scalar cosmological perturbations have been widely studied in fourth order gravity theories mainly in the metric formalism \cite{Perturbations_fR, Qstatic_Bean, delaCruzDombriz:2008cp}
originally developed for GR by Bardeen \cite{bardeen} as well as in the 1+3 covariant approach \cite{Perturbations_1+3}. 
%
Moreover, in the last years, attempts to numerically compute the matter power spectrum for classes of modified gravity theories by using several parameterisations, instead of solving the full set of equations have received increasing attention \cite{Codes}. Modifications to the linear order Einstein equations are thus introduced in terms of general functions of scale and time.
%
In general all these investigations proved that the perturbations growth depend on the scale (while in GR the evolution of dust matter is scale-independent \cite{Ananda} 
and that models claimed as viable are in fact almost ruled-out due to the matter power spectrum \cite{Dombriz_PRL,Dunsby_2013}.
Perturbations in scalar-tensor theories have been considered from different approaches, for example: the reconstruction problem was addressed in \cite{Gilles}, the density contrast evolution was studied in \cite{todas_density_contrast_ST_theories, Boisseau:2000pr, Amendola:2003wa} where in general, some intermediate approximations were performed,
%
the integrated Sachs-Wolfe effect was determined in \cite{Baccigalupi:2000je} and finally second order perturbations were presented in \cite{Tatekawa:2008bw}. In all the aforementioned studies where an equation for the evolution of the density contrast were sought, the study was developed in the JF (with the exception of \cite{Amendola:2003wa} which was carried in EF) and under the so-called quasistatic approximation. 
This approximation is performed disregarding all the time derivative terms for the gravitational potentials in the first order perturbed equations. In the context of modified gravity theories, this approximation has been considered as too aggressive \cite{Qstatic_Bean, delaCruzDombriz:2008cp} since neglecting time derivatives may remove important information preventing it from encapsulating all the features that correctly describe the perturbations evolution.

In this work we shall focus in the aforementioned type of scalar-tensor theories, the JFBD theories. Also, we shall work in the EF.
The following transformations \cite{Brans:2005ra,Olive,Gilles}
\begin{eqnarray}
\omega_0=-\frac{3-\alpha^{-2}}{2}\;\;;\;\;
\varphi=e^{-2\alpha\varphi_*}=A(\varphi_*)^{-2},
\end{eqnarray}
turn the action in the JF \eqref{actionJF} into the EF \cite{Brans:2005ra,Olive,Gilles}:
\begin{eqnarray}
%
S=\int\frac{{\rm d}^4x}{16\pi G_*}\sqrt{-g_*}\left[R_*-2\Lambda+2g^{\mu\nu}_*\partial_\mu\varphi_*\partial_\nu\varphi_*\right]\nonumber\\
+S_M[e^{2\alpha\varphi_*}g_{\mu\nu}^*;\psi]\,,
\label{actionEF}
\end{eqnarray}
where we have added a cosmological constant $\Lambda$. Note that in this frame the gravitational constant $G_*$ is not dependent on the scalar field. 
However, the matter term is now coupled directly to the scalar field through the coupling parameter $\alpha$. 
This coupling would mean that test particles do not follow geodesics, nor the inertial mass is conserved in this frame \cite{Brans:2005ra}.

In the present work, we address the problem of determining the exact equation for the evolution of matter density perturbations for JFBD theories in the EF.
The quasistatic limit of this equation was
 was obtained in \cite{Boisseau:2000pr} in the JF and in \cite{Amendola:2003wa} in the EF, where in both the standard quasistatic {\it a priori} approximations were performed.
This equation is similar to that in GR but instead of the bare gravitational constant $G_*$ the quasistatic equation appears with an effective gravitational constant, which depends on the model through the parameter $\omega_0$ or $\alpha$ (and in the JF on the scalar field also) but it is independent of the scale $k$ as in the GR case.

Once we obtain such an exact equation, we shall focus on obtaining solutions for the evolution of the matter density contrast with a $\Lambda\text{CDM}$ cosmological background supplemented by a JFBD field under certain viability constraints for the coupling. We shall then compare the results with those obtained in the standard $\Lambda\text{CDM}$ model, i.e., without any additional field.

\

This work is organised as follows:
In Section \ref{I.A}\textcolor{red}{-A} we revise the results for the Concordance $\Lambda\text{CDM}$ model as well as introducing the notation and several concepts.
We further deepen the discussion between the different frames in JFBD theories in Section \ref{I.B}\textcolor{red}{-B}.
In Section \ref{II} we introduce the background equations and obtain the perturbed equations of JFBD theories.
Section \ref{III} is devoted to solving the perturbed equations in order to obtain the evolution of the density contrast.
Then, in Section \ref{IV}, we particularise our results for several JFBD models.
In Section \ref{V} we summarise the conclusions of the present investigation.
Finally, in Appendix \ref{APPENDIX A} we show the coefficients of the general equation for the evolution of density perturbations.


\subsection{I-A. Density perturbations in $\Lambda$CDM}
\label{I.A}

Below, we summarise the results for density perturbations in the $\Lambda$CDM model as an introduction to the subject of density perturbations. The corresponding first order perturbed EE read
\begin{eqnarray}
\delta G^\mu_{\;\;\nu}=-8\pi G\delta T^\mu_{\;\;\nu},
\label{GRpert}
\end{eqnarray}
with $G^\mu_{\;\;\nu}$ the Einstein's tensor and $T^\mu_{\;\;\nu}$ the stress-energy tensor for matter.
At this stage we need to consider the perturbed metric. In this work we are interested in obtaining the evolution for density perturbations so, we just need to consider scalar perturbations in the metric. We work in the longitudinal gauge and in conformal time, so the perturbed FLRW metric reads
\begin{equation}
{\rm d}s^2\,=\,a^2(\eta)[(1+2\Phi){\rm d}\eta^2-(1-2\Psi)({\rm d}r^2+r^2{\rm d}\Omega_{2}^2)],
\label{metricpert}
\end{equation}
where $\Phi\equiv\Phi(\eta,\vec{x}),\;\Psi\equiv\Psi(\eta,\vec{x})$ are the Bardeen's potentials \cite{Bardeen}. Once the metric has been introduced, we can obtain the first order perturbed Christoffel symbols as well as the Einstein tensor in the longitudinal gauge \cite{Theor-tools}\footnote{Note that in \cite{Theor-tools}  the authors use another convention for the Riemann tensor so, a change needs to be performed, but the Christoffel symbols remain unchanged.}.
If we restrict ourselves to adiabatic perturbations and barotropic fluids we conclude that the perturbed stress-energy tensor can be expressed as 
\begin{eqnarray}
\delta T^{0}_{\;\;0}\,&=&\,\delta\rho = \rho_{0} \delta,\;\; \delta T^{i}_{\;\;j}\,=\, -\delta P \delta^{i}_{\;\;j} = -c_s^2 \delta^{i}_{\;\;j}\rho_{0}\delta,\nonumber\\
\delta T^{0}_{\;\;i}\,&=&\,-\delta T^{i}_{\;\;0}\,=-\,(1 + c_s^2)\rho_{0}\partial_{i}v,
\label{energymomentumpert}
\end{eqnarray}
where $\rho_0$ is the unperturbed average cosmological energy density for a fluid and $\rho$ is the perturbed energy density of the same cosmological fluid, $\delta$
is the density contrast, $c_s$ is the speed of sound and $v$ holds for the potential for velocity perturbations.
In the following, we shall assume that both perturbed and unperturbed matter obey the same equation of state, i.e., $\delta P/\delta \rho \equiv c_s^2 \equiv P_0/\rho_0$.
From \eqref{GRpert} and the corresponding stress-energy tensor conservation 
and after using the definitions above, we can obtain a second order differential equation for $\delta$. In particular for dust matter, i.e. $c_s^2=0$, it reads \cite{Alvaro's-thesis}
\begin{eqnarray}
\lefteqn{
\delta''+\mathcal{H}\frac{k^4-6\tilde{\rho}k^2-18\tilde{\rho}^2}{k^4-\tilde{\rho}(3k^2+9\mathcal{H}^2)}\delta'}\nonumber\\ &-\tilde{\rho}&\frac{k^4+9\tilde{\rho}(2\tilde{\rho}-3\mathcal{H}^2)-k^2(9\tilde{\rho}-3\mathcal{H}^2)}{k^4-\tilde{\rho}(3k^2+9\mathcal{H}^2)}\delta=0,
\label{ecdeltasRG}
\end{eqnarray}
where prime denotes derivative respect to conformal time $\eta$, $\tilde{\rho}\equiv4\pi G a^2\rho_0$ and $\mathcal{H}=a'/a$.
In the extreme sub-Hubble limit, i.e., $k>>\mathcal{H}$, \eqref{ecdeltasRG} becomes
\begin{eqnarray}
\delta''+\mathcal{H}\delta'-\tilde{\rho}\delta=0.
\label{ecdeltasRGsubHubble}
\end{eqnarray}
In the case of GR, the sub-Hubble limit matches the quasistatic limit ($k\rightarrow\infty$) because the sub-Hubble approximation is not dependent on the mode $k$.

%

\subsection{I-B. Discussion on the transformation between frames}
\label{I.B}

Despite presenting our results in the EF we shall compare them with results obtained in the JF at the end of Section \ref{III}. Therefore the quantities $\delta_*$ and $g_{\mu\nu}^*$, defined in the EF need to be related with their counterparts $\delta$ and $g_{\mu\nu}$, defined in the JF. The relations between the aforementioned quantities is as follows
\cite{Olive,Gilles}:
\begin{eqnarray}
a_*(\eta_*)\,=\,A(\varphi_*)^{-1}a(\eta),\;\; 
\rho_*\,=\,A(\varphi_*)^4\rho,
\label{changeEFJF}
\end{eqnarray}
where $A(\varphi_*) \equiv e^{\alpha\varphi_*}$ is related to the coupling of the scalar field and the metric in the EF. The conformal time $\eta$ is the same in both frames ($\eta_*\equiv\eta$).

Taking into account these relations, the density contrast and Hubble parameter change as follows
\begin{eqnarray}
\delta_*=\frac{\delta(\rho_*)}{\rho_{0*}}=\delta+4\alpha\,\delta\varphi_*,\nonumber\\
\mathcal{H}_*=\frac{a'_*}{a_*}=\mathcal{H}-\frac{\dot{A}}{A}\,\varphi'_{*}=\mathcal{H}-\alpha\varphi'_*,
\label{changedeltaEFJF}
\end{eqnarray}
where $\dot{ }\equiv\partial_\varphi$. From now on, let us drop the notation (*) for EF quantities, since we shall work throughout the manuscript in this frame. 



\begin{figure*} [htbp] 
\centering
  	\includegraphics[width=0.420\textwidth]{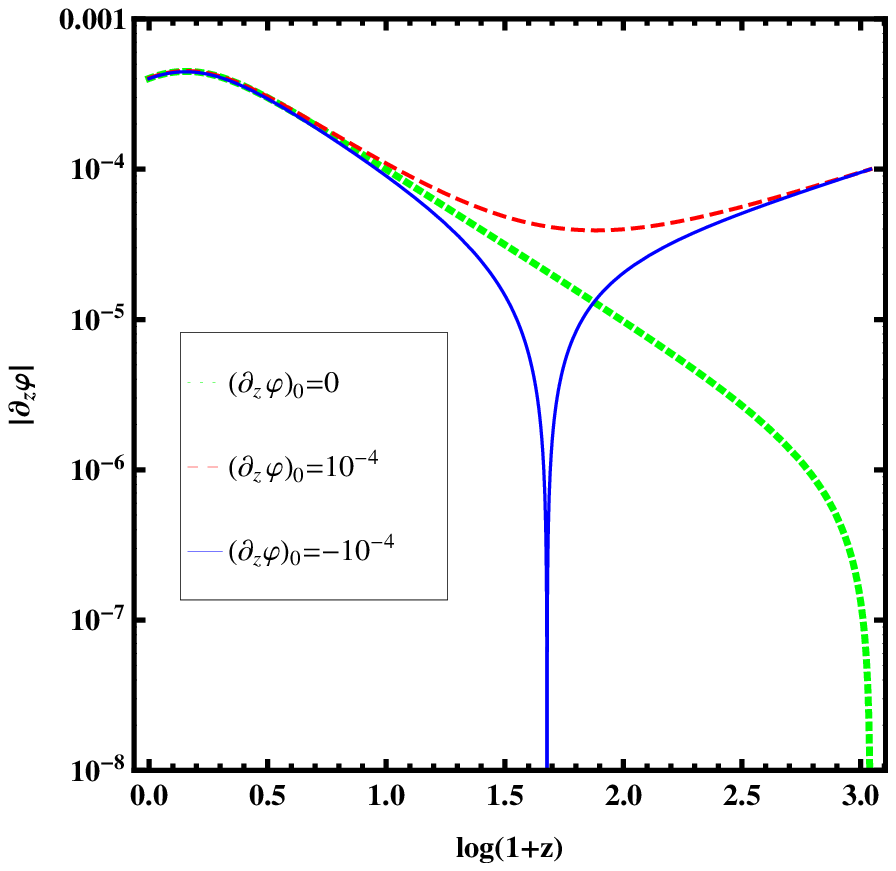}
		\includegraphics[width=0.450\textwidth]{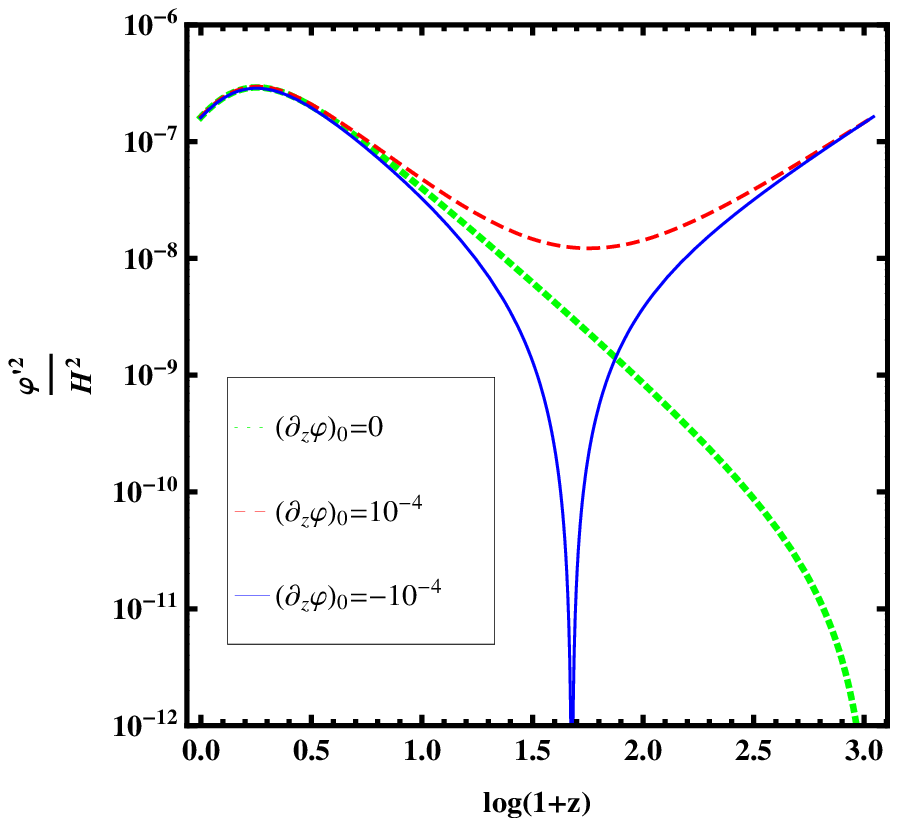}
		\caption{\footnotesize{Model $\alpha=10^{-3}$ for initial conditions $\partial_z\varphi = (-10^{-4},0,10^{-4})$. Left panel: Redshift evolution of the scalar field derivative. Right panel: Comparison of $\varphi'^2$ -- note that prime denotes conformal time derivative -- with $\mathcal{H}^2$. We conclude that the requirement of a negligible stress-energy tensor of the scalar field is satisfied regardless of the initial conditions for $\partial_z\varphi$.}}
  \label{scalarfieldsC1-1000}
\end{figure*}

\section{JFBD theories}
\label{II}

\subsection{Background evolution in JFBD theories}

The corresponding modified EE 
obtained from the action \eqref{actionEF} read:
\begin{eqnarray}
G_{\mu\nu}+[2V(\varphi)-g^{\alpha\beta}\nabla_\alpha\varphi\nabla_\beta\varphi]g_{\mu\nu}+2\nabla_\mu\varphi\nabla_\nu\varphi
\nonumber\\
=-8\pi G T_{\mu\nu},
\label{g_eqn}
\end{eqnarray}
and the equation of motion for the scalar field yields:
\begin{eqnarray}
\square\varphi\,=\,\nabla^\alpha\nabla_\alpha\varphi\,=\,-4\pi G\alpha T,
\label{scalar_eqn}
\end{eqnarray}
where $\alpha=const$ defines the coupling between the metric and the scalar field.
In addition, if we take the trace of the covariant derivative in \eqref{g_eqn} and use \eqref{scalar_eqn} the conservation equation becomes:
\begin{eqnarray}
\nabla_\mu T^\mu_{\;\nu}\,=\,\alpha T \partial_\nu\varphi.
\label{cons_eqn}
\end{eqnarray}

For a spatially flat FLRW metric the components $(00)$ and $(rr)$ of  \eqref{g_eqn} with $\varphi=\varphi(\eta)$ and a perfect fluid read \cite{Olive,Gilles}
\begin{eqnarray}
&&3\mathcal{H}^2
-\varphi'^2\,=\,2\tilde{\rho} +\Lambda a^2,
\label{g_comps_0}\\
&&\mathcal{H}^2-\mathcal{H}'-\varphi'^2\,=\,  \tilde{\rho}(1+c_s^2),
\label{g_comps_j}
\end{eqnarray}
%
Then, equation \eqref{scalar_eqn} for the FLRW metric background gives
\begin{eqnarray}
\varphi''+2\mathcal{H}\varphi'\,=\,- \tilde{\rho}(1-3c_s^2)\alpha ,
\label{scalar_comps}
\end{eqnarray}
and for the temporal component of the conservation equation \eqref{cons_eqn} we get:
\begin{eqnarray}
\frac{\rho_0'}{\rho_0}=-3\mathcal{H}(1+c_s^2)+\alpha(1-3c_s^2)\varphi'.
\label{cons_comps}
\end{eqnarray}
In summary we are left with four background equations, two from the gravitational field \eqref{g_comps_0}-\eqref{g_comps_j}, one from the scalar field \eqref{scalar_comps} and one from the conservation equation \eqref{cons_comps}.


\subsection{Perturbations in JFBD theories}

The first order perturbed equations of the gravitational field in JFBD theories
\eqref{g_eqn} become
\begin{eqnarray}
&&\delta G^\mu_\nu-\nabla_\alpha\varphi\nabla_\beta\varphi \delta^\mu_\nu(\delta g^{\alpha\beta})+2\nabla_\alpha\varphi\nabla_\nu\varphi(\delta g^{\mu\alpha})\nonumber\\
&&+\,2\left(
-\delta^\mu_\nu g^{\alpha\beta}\nabla_\beta\varphi\nabla_\alpha +g^{\alpha\mu}\nabla_\nu\varphi\nabla_\alpha+\nabla^\mu\varphi\nabla_\nu\right)\delta\varphi\nonumber\\
&&=-8\pi G\, \delta T^\mu_\nu.
\label{EEs-c}
\end{eqnarray}
Using the perturbed metric \eqref{metricpert}, the perturbed stress-energy tensor \eqref{energymomentumpert} and assuming that the background equations hold, the components of equation \eqref{EEs-c} $(00)$, $(ii)$, $(0i)\equiv(i0)$ and $(ij)$, where $i,j=1,2,3$, $i\neq j$ in Fourier space, respectively read,
\begin{eqnarray}
&&-k^2\Psi-3\mathcal{H}(\Psi'+\mathcal{H}\Phi)+\Phi\varphi'^2-\varphi'\delta\varphi'
=\tilde{\rho}\delta,\\
\nonumber\\
&&\Psi''+2(\mathcal{H}^2+2\mathcal{H}')\Phi+2\mathcal{H}(\Phi'+2\Psi')-k^2(\Phi-\Psi)\nonumber\\
&&+\,2\Phi\varphi'^2
 - 2\varphi'\delta\varphi'=2\tilde{\rho}c_s^2\delta,\\
 \nonumber\\
&&\Psi'+\mathcal{H}\Phi-\varphi'\delta\varphi=-\tilde{\rho}(1+c_s^2)v,\\
\nonumber\\
&&\partial_i\partial_j(\Psi-\Phi)=0.
\label{equality}
\end{eqnarray}
The last equation \eqref{equality} implies that both scalar potentials $\Phi$ and $\Psi$ are equal as in GR, so the previous equations, for $(00)$, $(ii)$ and $(0i)\equiv (i0)$ components can be rewritten respectively as
\begin{eqnarray}
&&3\mathcal{H}\Phi'+(k^2+3\mathcal{H}^2-\varphi'^2)\Phi+\varphi'\delta\varphi'
=-\tilde{\rho}\delta,
\label{ec00}\\
\nonumber\\
&&\Phi''+3\mathcal{H}\Phi'+(\mathcal{H}^2+2\mathcal{H}'+\varphi'^2)\Phi-\varphi'\delta\varphi'
=\tilde{\rho}c_s^2\delta,\nonumber\\
\label{ecii}
\\
&&\Phi'+\mathcal{H}\Phi-\varphi'\delta\varphi=-\tilde{\rho}(1+c_s^2)v.
\label{ec0i}
\end{eqnarray}
For the scalar field equation \eqref{scalar_eqn} the first order perturbed equation yields:
\begin{eqnarray}
&&\delta g^{\alpha\beta}\nabla_\alpha\nabla_\beta\varphi+g^{\alpha\beta}\left(\delta\varphi_{,\alpha\beta}-\delta\Gamma^\gamma_{\beta\alpha}\varphi_{,\gamma}-\Gamma^\gamma_{\beta\alpha}\delta\varphi_{,\gamma}\right)\nonumber\\
&&\,=\,-4\pi G\alpha\delta T,
\end{eqnarray}
which for the metric (\ref{metricpert}) and after using the fact that $\Phi = \Psi$  reads
\begin{eqnarray}
&&4\varphi'\Phi'+2(\varphi''+2\mathcal{H}\varphi')\Phi-\delta\varphi''-2\mathcal{H}\delta\varphi'
-
k^2
\delta\varphi\nonumber\\
&&\,=\,(1-3c_s^2)\alpha\tilde{\rho}\delta,
\label{ecesc}
\end{eqnarray}
Finally, the conservation equation \eqref{cons_eqn} up to first order in perturbations becomes:
\begin{eqnarray}
&&\partial_\mu\delta T^\mu_\nu+\delta\Gamma^\mu_{\alpha\mu}T^\alpha_\nu+\Gamma^\mu_{\alpha\mu}\delta T^\alpha_\nu-\delta\Gamma^\alpha_{\mu\nu}T^\mu_\alpha
-\Gamma^\alpha_{\mu\nu}\delta T^\mu_\alpha\nonumber\\
&&\,=\,\alpha \delta T \partial_\nu\varphi+\alpha T\partial_\nu\delta\varphi,
\end{eqnarray}
whose temporal and spatial components using \eqref{cons_comps} and \eqref{equality} read respectively
\begin{eqnarray}
&&3\Phi'-\frac{\delta'}{1+c_s^2}+k^2v=-\frac{1-3c_s^2}{1+c_s^2}\alpha\,\delta\varphi',
\label{eccons0}\\
&&\Phi+v'+\left(1-3c_s^2\right)\left(\mathcal{H}+\alpha\varphi'\right)v+\frac{c_s^2}{1+c_s^2}\delta \nonumber\\
&&\,=\,-\frac{1-3c_s^2}{1+c_s^2}\alpha\,\delta\varphi.
\label{ecconsi}
\end{eqnarray}
%

\section{Evolution of density perturbations in JFBD theories}
\label{III}

The goal of this section consists of getting a differential equation for dust matter, i.e., $c_s^2=0$, density contrast $\delta$. We detail below the method we have followed to obtain such an equation.
First, we must identify which equations are independent from each other. Thus it turns out that equation \eqref{ecii} can be obtained by combining \eqref{ec00}, its first derivative with respect to time, \eqref{ec0i}, \eqref{ecesc} and \eqref{eccons0}. The same happens with equation \eqref{ecconsi}, which is dependent on some of the aforementioned equations and their derivatives. Therefore, we are left with a system of four equations, namely \eqref{ec00}, \eqref{ec0i}, \eqref{ecesc} and \eqref{eccons0}.
As usual $v$ can be obtained from  equation
\eqref{eccons0} in terms of other perturbed quantities
so we end up with a system of three equations and five variables $\left\{\Phi,\ \Phi',\ \delta\varphi,\ \delta\varphi',\ \delta\varphi''\right\}$. This forces us to differentiate and combine the aforementioned three equations \eqref{ec00}, \eqref{ec0i} and \eqref{ecesc} in order to get a differential
equation for $\delta$.

After substituting $v$ in the equations \eqref{ec00}, \eqref{ec0i} and \eqref{ecesc} we can combine them and algebraically isolate the variables $\left\{ \delta\varphi,\ \delta\varphi',\ \delta\varphi''\right\}$. Thus we obtain:
\begin{eqnarray}
\delta\varphi &=& \delta\varphi(\delta,\delta',\Phi,\Phi'), \nonumber\\
 \delta\varphi' &=& \delta\varphi'(\delta,\Phi,\Phi'),\\
\delta\varphi'' &=& \delta\varphi''(\delta,\delta',\Phi,\Phi'), \nonumber
\label{varphi_despejados}
\end{eqnarray}
where we can see that $\delta\varphi'$ does not depend on $\delta'$.
Differentiating $\delta\varphi'$ and equating the result with $\delta\varphi''$ we can solve for $\Phi''$ and obtain
\begin{eqnarray}
\Phi'' = \Phi''(\delta,\Phi,\Phi')\,,
\label{Phi2_casi_final}
\end{eqnarray}
since the $\delta'$ coefficients turn out to cancel.

At this stage, we need to obtain $\Phi$ and $\Phi'$ in terms of $\delta$ and its derivatives and hence we need two further equations. We obtain them by combining equations \eqref{ec00} and \eqref{ec0i} in such a way that $\Phi'$ cancels. Then we differentiate the obtained equation and substitute the variables we already know from
\eqref{varphi_despejados} and \eqref{varphi_despejados}. Thus, from this combination we get:
\begin{eqnarray}
\mathcal{Q}_0(\delta,\delta',\delta'',\Phi,\Phi') = 0.
\label{ecQ0}
\end{eqnarray}

Furthermore, we can differentiate this last equation again to obtain:
\begin{eqnarray}
\mathcal{Q}_1(\delta,\delta',\delta'',\delta''',\Phi,\Phi') = 0,
\label{ecQ1}
\end{eqnarray}
where we have used the already obtained expression for $\Phi''$ given in \eqref{Phi2_casi_final}.
From last two equations \eqref{ecQ0} and \eqref{ecQ1} we can obtain algebraically $\Phi$ and $\Phi'$:
\begin{eqnarray}
\Phi = \Phi(\delta,\delta',\delta'',\delta'''),\
\Phi' = \Phi'(\delta,\delta',\delta'',\delta''').
\label{final}
\end{eqnarray}

The last step just consists of differentiating $\Phi$ in \eqref{final} and equate the result with $\Phi'$ also given in \eqref{final}. This way, a fourth order differential equation for $\delta$ is obtained:
\begin{eqnarray}
\mathcal{C}_4\delta^{iv}+\mathcal{C}_3\delta'''+\mathcal{C}_2\delta''+\mathcal{C}_1\delta'+\mathcal{C}_0\delta=0.
\label{ecdeltagen}
\end{eqnarray}
The actual coefficients $\mathcal{C}_i$ $(i=0,1,2,3,4)$ are shown in Appendix \ref{APPENDIX A}.
\
As a consistency check, we remark at this stage that the well-known second order differential equation for GR \eqref{ecdeltasRG}
is recovered provided that  the scalar field and its derivatives tend to zero by a careful procedure.  Note also that the above described procedure is completely general to first order for scalar perturbations in metric formalism for arbitrarily general scalar-tensor theories such as those including potential or non-constant coupling $\alpha$.

If one is now interested in studying sub-Hubble modes, we can perform the sub-Hubble approximation ($k>>\mathcal{H}$)
in \eqref{ecdeltagen}. In order to do so, let us define the parameter $\epsilon\equiv\mathcal{H}/k$ that allows us to perform a perturbative expansion on the $\mathcal{C}_i$ coefficients and keep only the leading terms in $\epsilon$. Once the approximation is performed, \eqref{ecdeltagen} reads
\begin{eqnarray}
&&\mathcal{C}_{4,2}\delta^{iv}+\mathcal{C}_{3,2}\delta'''+(\mathcal{C}_{2,0}+\mathcal{C}_{2,2})\delta''\nonumber\\
&&\,+\,(\mathcal{C}_{1,0}+\mathcal{C}_{1,2})\delta'+(\mathcal{C}_{0,0}+\mathcal{C}_{0,2})\delta=0,
\label{ecdeltaSH}
\end{eqnarray}
where the coefficients $\mathcal{C}_{i,j}$ are also given in Appendix \ref{APPENDIX A}.

\begin{figure*} [htbp] 
	\centering
		\includegraphics[width=0.329\textwidth]{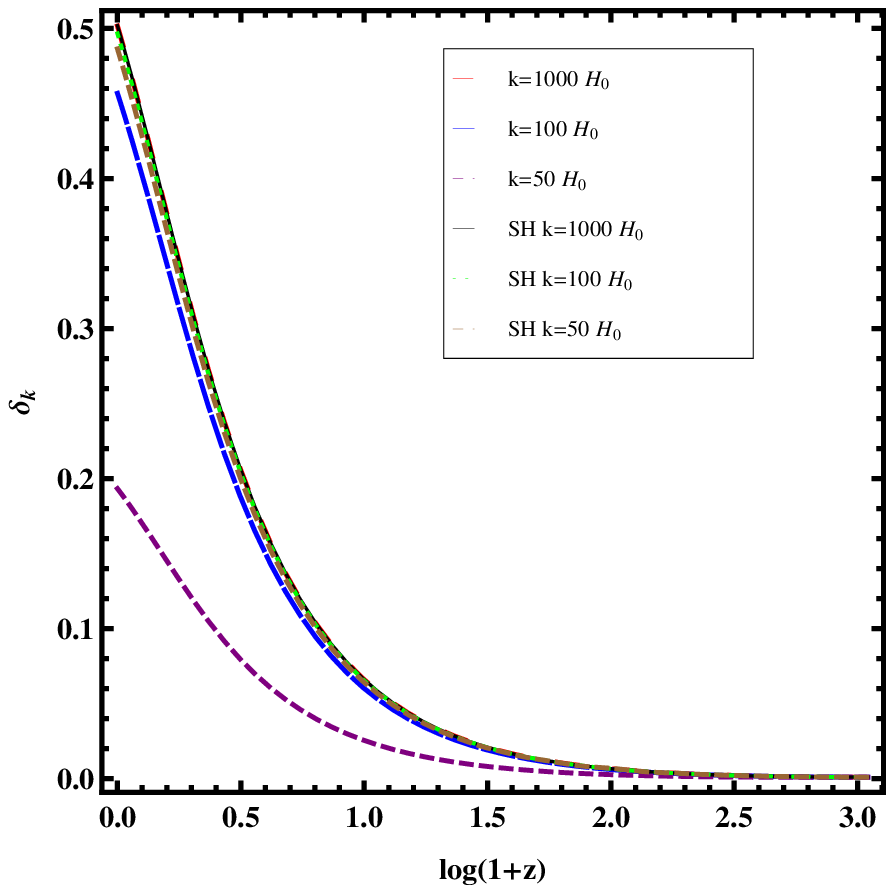}
		\includegraphics[width=0.329\textwidth]{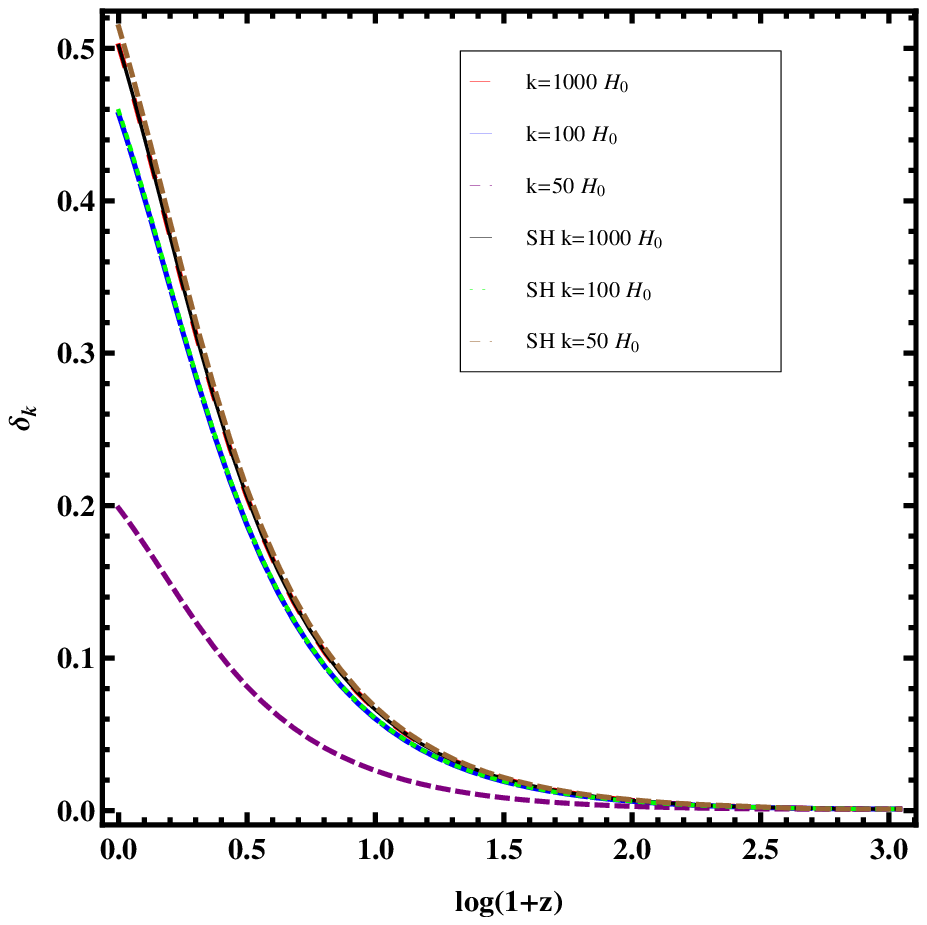}
		\includegraphics[width=0.329\textwidth]{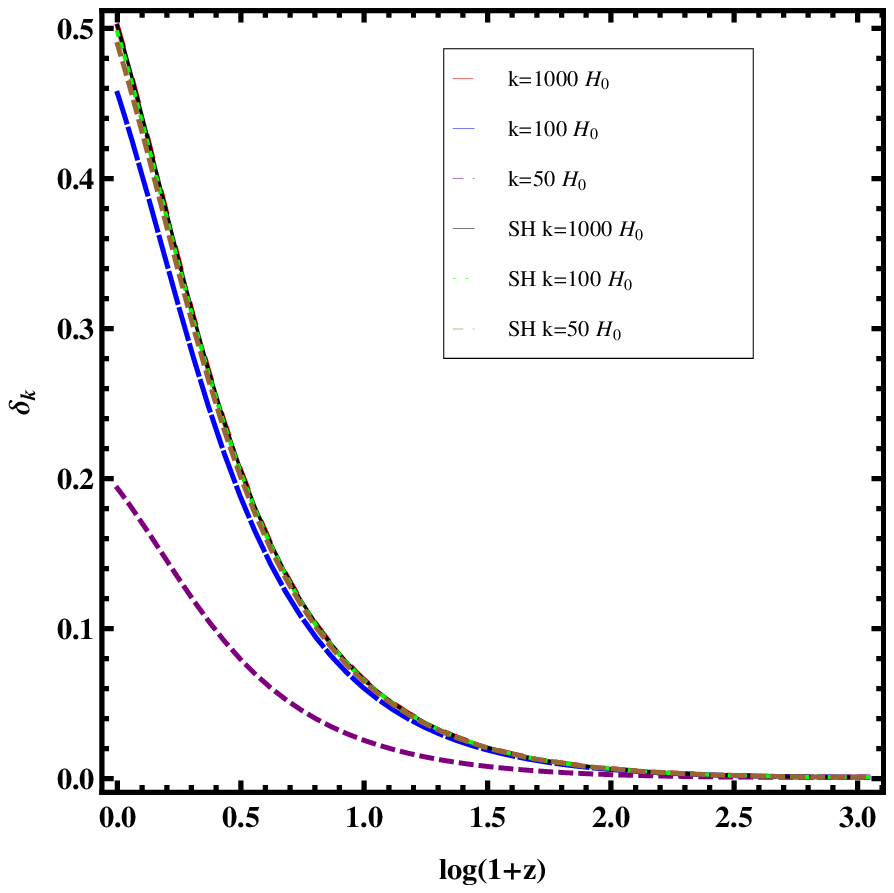}
 \caption{\footnotesize{Model $\alpha=10^{-3}$: $\delta_k$ evolution for $\partial_z\varphi_0=10^{-4}$ (left), $\partial_z\varphi_0=0$ (centre) and $\partial_z\varphi_0=-10^{-4}$ (right), for the modes $k\;=\;50\mathcal{H}_0,100\mathcal{H}_0,1000\mathcal{H}_0$. Both
 general equation \eqref{ecdeltagen} and sub-Hubble approximation \eqref{ecdeltaSH} have been plotted in the redshift range from 1100 to 0.
 We can see in the centre panel that the sub-Hubble approximation is an upper limit for every specific $k$ but that it depends on $k$ in contrast with the $\Lambda$CDM case. $\delta(z=1100)=10^{-3}$ and $\delta'(z=1100)=0$ have been used as initial conditions.}}
 \label{C1-1000delta}
\end{figure*}

\begin{figure*} [htbp] 
	\centering
		\includegraphics[width=0.329\textwidth]{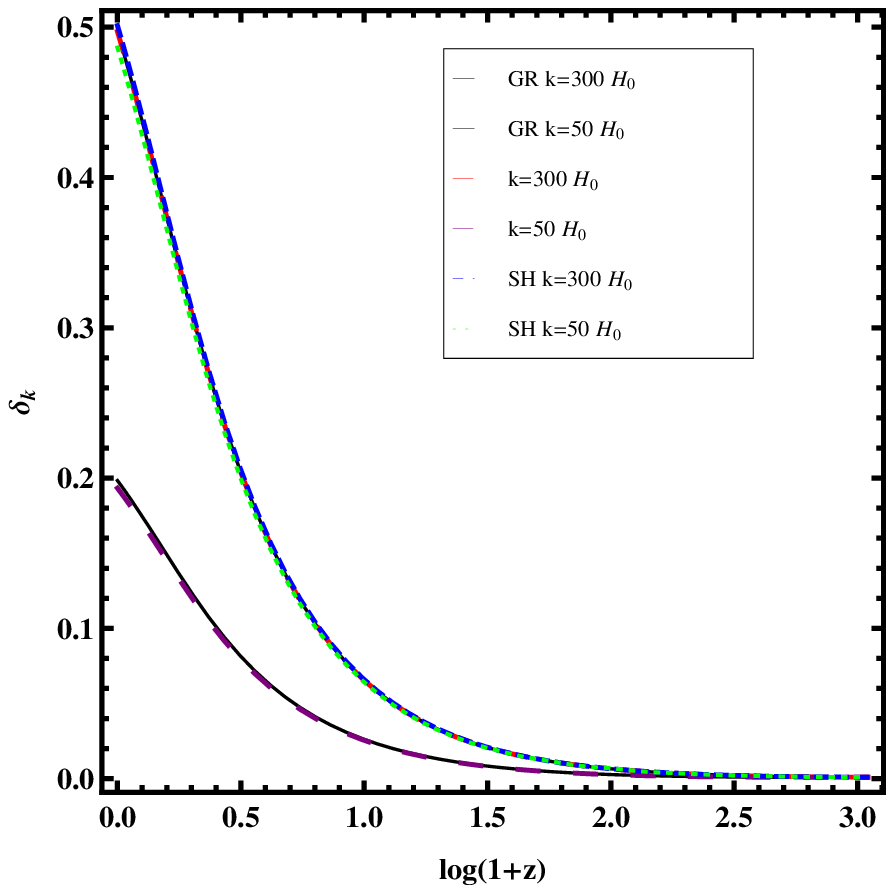}
		\includegraphics[width=0.329\textwidth]{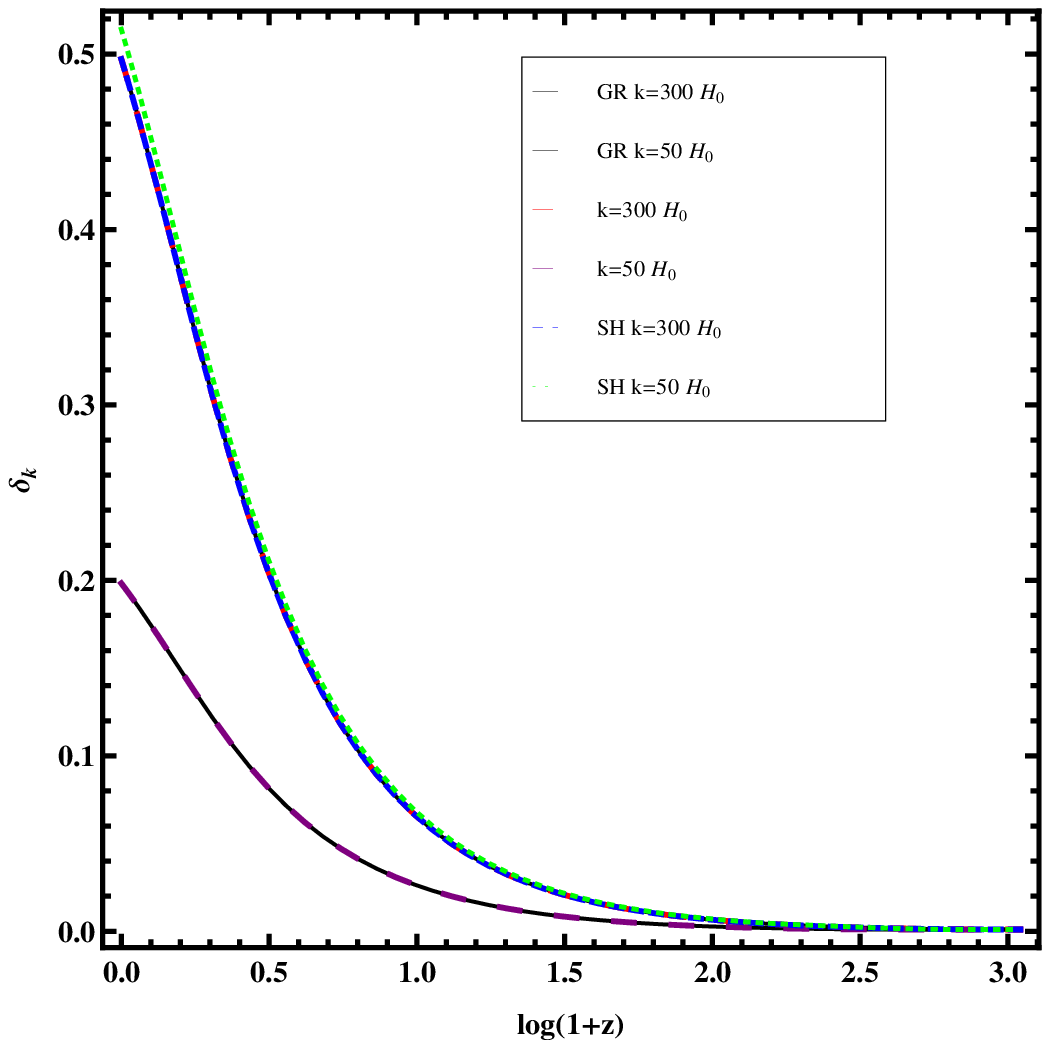}
		\includegraphics[width=0.329\textwidth]{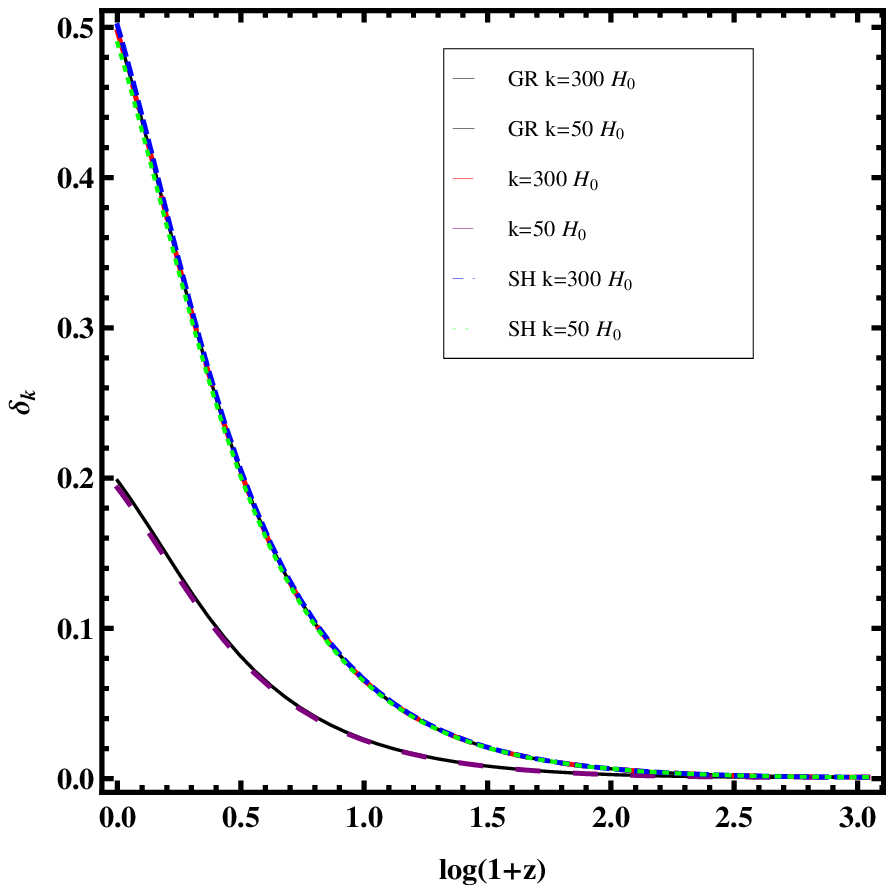}
 \caption{\footnotesize{Model $\alpha=10^{-3}$: $\delta_k$ for $\partial_z\varphi_0=10^{-4}$ (left), $\partial_z\varphi_0=0$ (centre) and $\partial_z\varphi_0=-10^{-4}$ (right), for the modes $k\;=\;50\mathcal{H}_0$ and $300\mathcal{H}_0$ described by the general equation \eqref{ecdeltagen}, sub-Hubble approximation \eqref{ecdeltaSH} and $\Lambda$CDM \eqref{ecdeltasRG} in the redshift range from 1100 to 0. $\delta(z=1100)=10^{-3}$ and $\delta'(z=1100)=0$ have been used as initial conditions. We can see that $\Lambda$CDM and the general equation are very similar and that the sub-Hubble approximation represents a good approximation for the highest mode $k\,=\,300\mathcal{H}_0$.}}
 \label{C1-1000deltakcomp}
\end{figure*}

Furthermore, the quasistatic approximation, which at this point is equivalent to make ($k\rightarrow\infty$ or $\epsilon\rightarrow0$) in \eqref{ecdeltagen} or \eqref{ecdeltaSH}, reads
\begin{eqnarray}
\delta''+(\mathcal{H}+\alpha\varphi')\delta'-(1+\alpha^2)\tilde{\rho}\delta=0.
\label{ecdeltaQSA}
\end{eqnarray}
In this case the equation becomes second order in $\delta$, since the $\delta^{iv}$ and $\delta'''$ coefficients comprise a power of $k^2$ less than
the coefficients corresponding to $\delta'',\delta',\delta$ and therefore are severely suppressed. Now by using
equations \eqref{changeEFJF} and \eqref{changedeltaEFJF}, the equation \eqref{ecdeltaQSA} can be translated into the JF yielding
\begin{eqnarray}
\delta''+\mathcal{H}\delta'-4\pi G_*A^2(1+\alpha^2)\rho\delta=0.
\label{ecdeltaQSAJF}
\end{eqnarray}
Note that last equation involves the quantities $\delta$, $\mathcal{H}$ and $\rho$ written in the JF.
Here we can see how in the equation for the EF \eqref{ecdeltaQSA} the gravitational constant $G$ (included in the $\tilde{\rho}$ definition) does not depend on the scalar field (even though it does on the theory through $\alpha$) while in the equation for the JF \eqref{ecdeltaQSAJF} the gravitational constant depends on $A=A(\varphi)$.

Finally let us mention that the quasistatic approximation has been obtained in the EF, as given in \eqref{ecdeltaQSA}, in \cite{Amendola:2003wa} and in the JF, as given in \eqref{ecdeltaQSAJF}, in 
 \cite{Boisseau:2000pr}. We present this result as a consistency check for the more general equations \eqref{ecdeltagen} and \eqref{ecdeltaSH}.


\begin{figure*} [htbp] 
	\centering
		\includegraphics[width=0.329\textwidth]{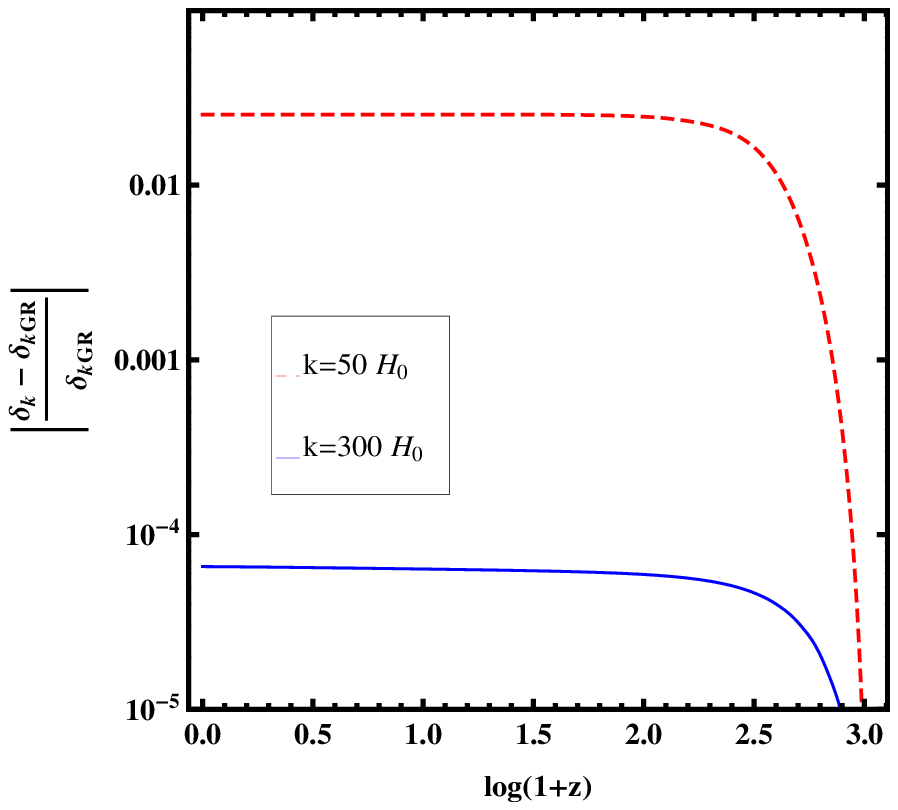}
		\includegraphics[width=0.329\textwidth]{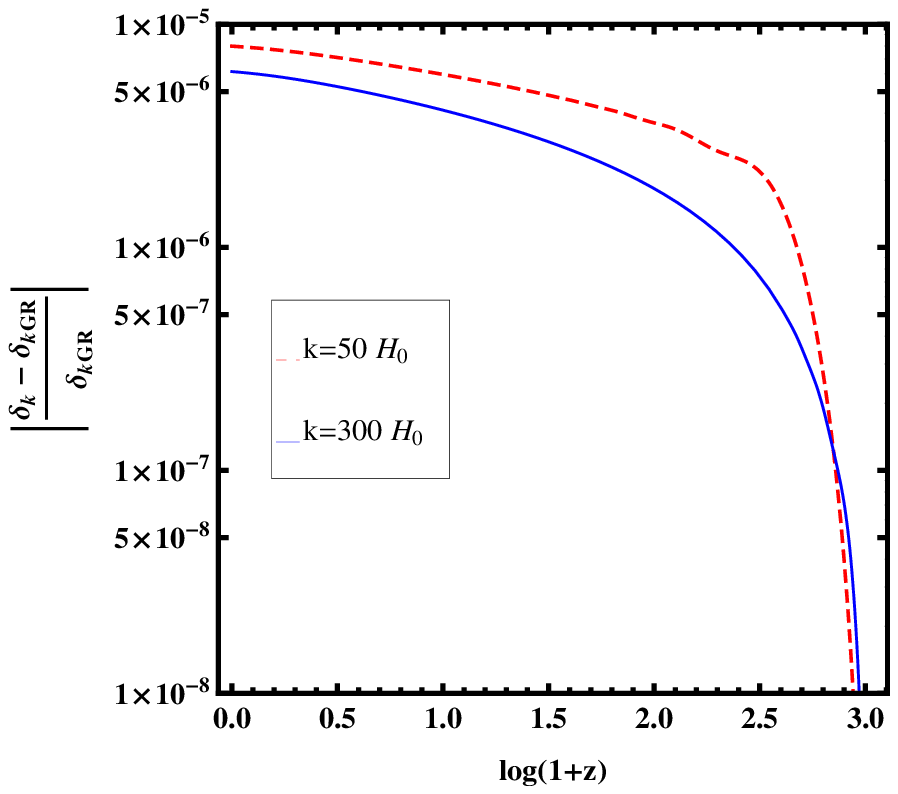}
		\includegraphics[width=0.329\textwidth]{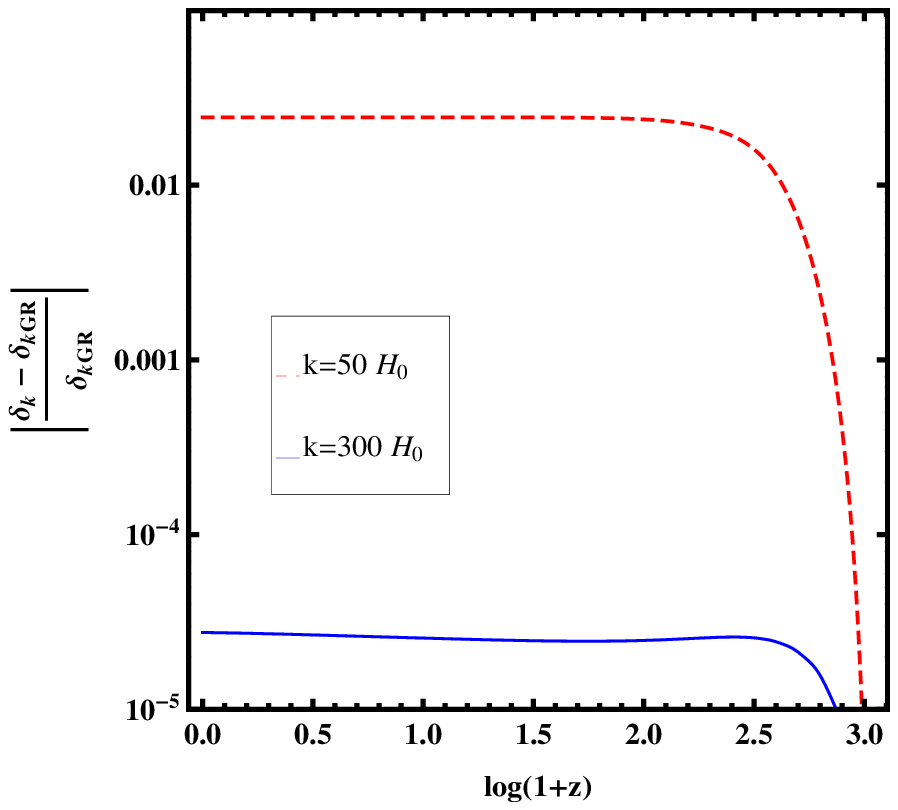}
 \caption{\footnotesize{Model $\alpha=10^{-3}$: Predicted separation of $\delta_k$ for $\partial_z\varphi_0=10^{-4}$ (left), $\partial_z\varphi_0=0$ (centre) and $\partial_z\varphi_0=-10^{-4}$ (right) described by the general equation \eqref{ecdeltagen} with respect to $\lambda$CDM solution. The studied modes were
$k\;=\;50\mathcal{H}_0,\,300\mathcal{H}_0$ in the redshift range from 1100 to 0. $\delta(z=1100)=10^{-3}$ and $\delta'(z=1100)=0$ have been used as initial conditions. As we can see, the difference with $\Lambda$CDM increases when $\partial_z\varphi_0 \neq 0$ and for the smallest mode.
}}
 \label{C1-1000deltaerror}
\end{figure*}

\section{Results}
\label{IV}

In order to solve the equations obtained in the previous section and describe the evolution of the density contrast $\delta$ let us consider the scalar field $\varphi$ contribution to the background to be negligible. Thus, the cosmological background would coincide with the $\Lambda$CDM solution, where Friedmann equation for the
spatially flat and dust plus cosmological constant mixture universe
becomes $H^2(z)=H^2_0\left[\Omega_M(1+z)^3+\Omega_\Lambda\right]$ with $\Omega_\Lambda=1-\Omega_M$ and $\Omega_M\sim0.3$ \cite{Ade:2013zuv}.

In addition, as we shall present our results as a function of redshift, we are required to translate our equations
from conformal time to cosmic time and accordingly we are able to use the analytic solution for the cosmological factor
as given in the $\Lambda$CDM model for dust (see for instance \cite{Alvaro's-thesis})
\begin{eqnarray}
a(t)=\left(\frac{1-\Omega_\Lambda}{\Omega_{\Lambda}}\right)^{1/3}\sinh^{(2/3)}\left(\frac{3\sqrt{\Omega_\Lambda}}{2}H_0 t\right),
\label{aSol}
\end{eqnarray}
We still need to compute the background evolution of the scalar field, which as we can see in equation \eqref{scalar_comps} is second order in $\varphi$ so it requires two initial conditions. We shall consider $\varphi(z=1100)=0$ as the scalar field initial value while we shall take the values $\left.\frac{d\varphi}{dz}\right|_{z=1100}=(-10^{-4},\ 0,\ 10^{-4})$ for the derivative with respect to the redshift of the scalar field.

We also need to specify the value of the coupling $\alpha$. The values under study shall be $\alpha = 10^{-3}$ and $\alpha = 10^{-1}$. The first value is chosen
because it constitutes an upper limit for the currently valid values \cite{Bertotti}\footnote{More precisely, the actual upper limit stablished by the Cassini spacecraft for $\alpha$ is $\alpha<3.54\cdot10^{-3}$ \cite{Bertotti}.} whereas the second value is chosen
for a better understanding on the $\alpha$ parameter dependence. Once the evolution of the scalar field is computed, we shall check that it is indeed negligible when compared with standard matter components and that therefore the aforementioned assumption is well founded.

Now that the background is totally specified, we can solve equations \eqref{ecdeltagen}, \eqref{ecdeltaSH} and \eqref{ecdeltaQSA} and compare
the evolution with the $\Lambda$CDM predictions as given by  \eqref{ecdeltasRG}. As \eqref{ecdeltagen} and \eqref{ecdeltaSH} are fourth order equations, they require four initial conditions for $\delta$. We use $\delta(z=1100)=10^{-3}$ and $\delta'(z=1100)=0$ as arbitrary conditions of the model and we obtain the conditions for the second and third derivatives from the GR equation \eqref{ecdeltasRG} and its first derivative.


\subsection{Case $\alpha=10^{-3}$}

For this case, the evolution of the background scalar field $\varphi$ is represented in Figure \ref{scalarfieldsC1-1000} for several initial conditions. It can be seen in the left panel that at recent epochs the first derivatives of the scalar field are undistinguishable from each other, but for the negative initial condition, $\partial_z\varphi$ crosses zero at some redshift. Right panel shows that the scalar field stress-energy tensor is smaller than the $\Lambda$CDM counterpart. Consequently our approximations for the background are fully justified in this case.

We show in Figures \ref{C1-1000delta} and \ref{C1-1000deltakcomp} the evolution of the scalar perturbations for several $k-$modes. Figure \ref{C1-1000delta} shows the evolution of several modes described by the general equation \eqref{ecdeltagen} and the sub-Hubble approximation \eqref{ecdeltaSH}. We can observe how the sub-Hubble approximation is able to describe the evolution for the biggest $k-$modes but it fails when it is dealing with smaller modes. Figure \ref{C1-1000deltakcomp} shows the evolution for modes $k = 50\mathcal{H}_{0},\, 300\mathcal{H}_0$ ($\mathcal{H}_0$ denotes the Hubble parameter evaluated today) in the general equation and the sub-Hubble approximation in addition to the evolution as predicted by the $\Lambda$CDM model. We can observe a slight deviation from the $\Lambda$CDM evolution in the models with non-zero initial condition for $\varphi'$ and for modes with $k\sim50\mathcal{H}_0$.

Additionally, given that the evolution of perturbations is similar to that of $\Lambda$CDM, we compare in Figure \ref{C1-1000deltaerror}  the relative error. In this figure we plot
the evolution of two modes ($k = 50\mathcal{H}_{0},\, 300\mathcal{H}_0$)
and observe that the difference with $\Lambda$CDM is bigger for the smallest mode ($k = 50\mathcal{H}_0$) for the cases with non-zero initial condition for $\varphi'$. Furthermore, we observe that the difference increases monotonously with the redshift for both modes.

\begin{figure*} [htbp] 
	\centering
	\includegraphics[width=0.309\textwidth]{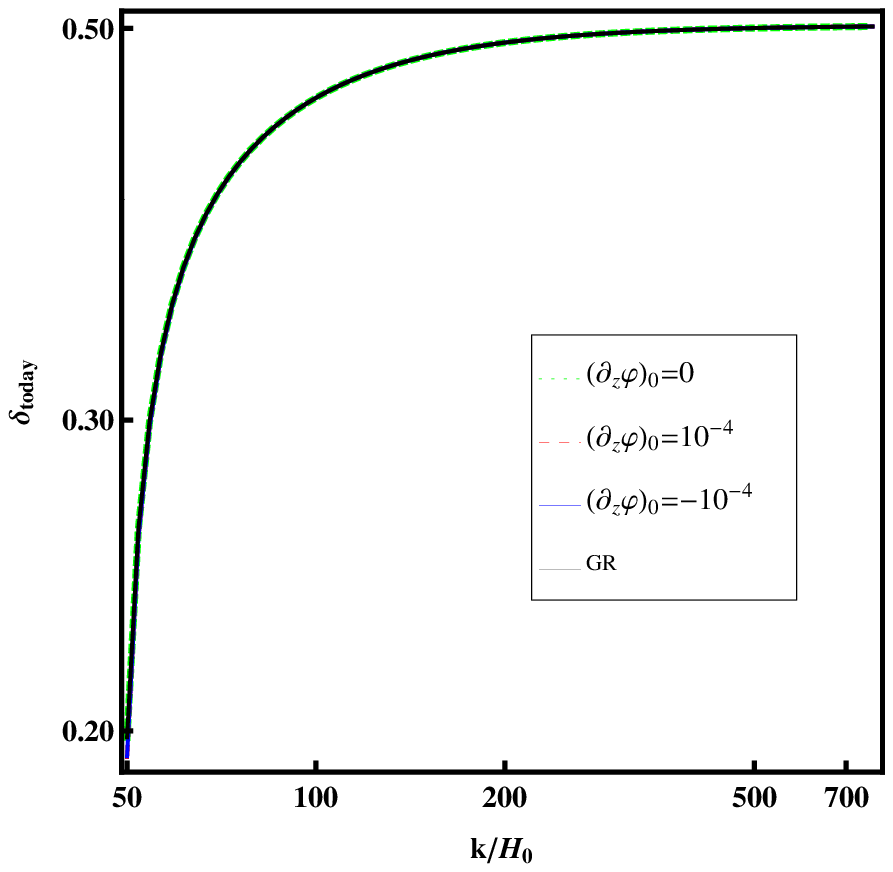}
	\includegraphics[width=0.329\textwidth]{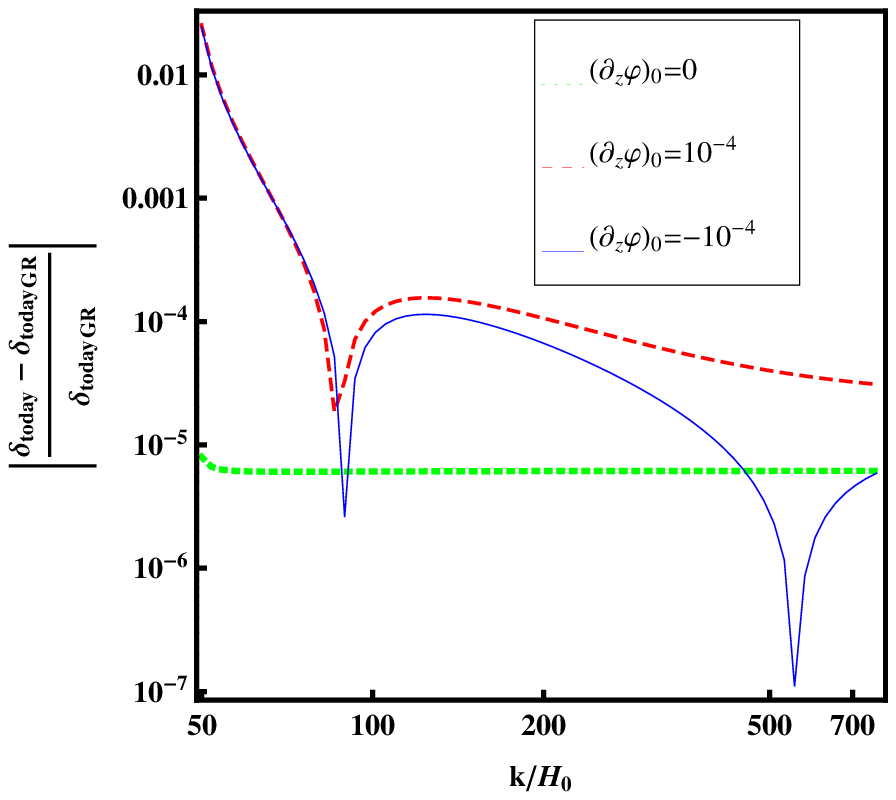}
	\includegraphics[width=0.329\textwidth]{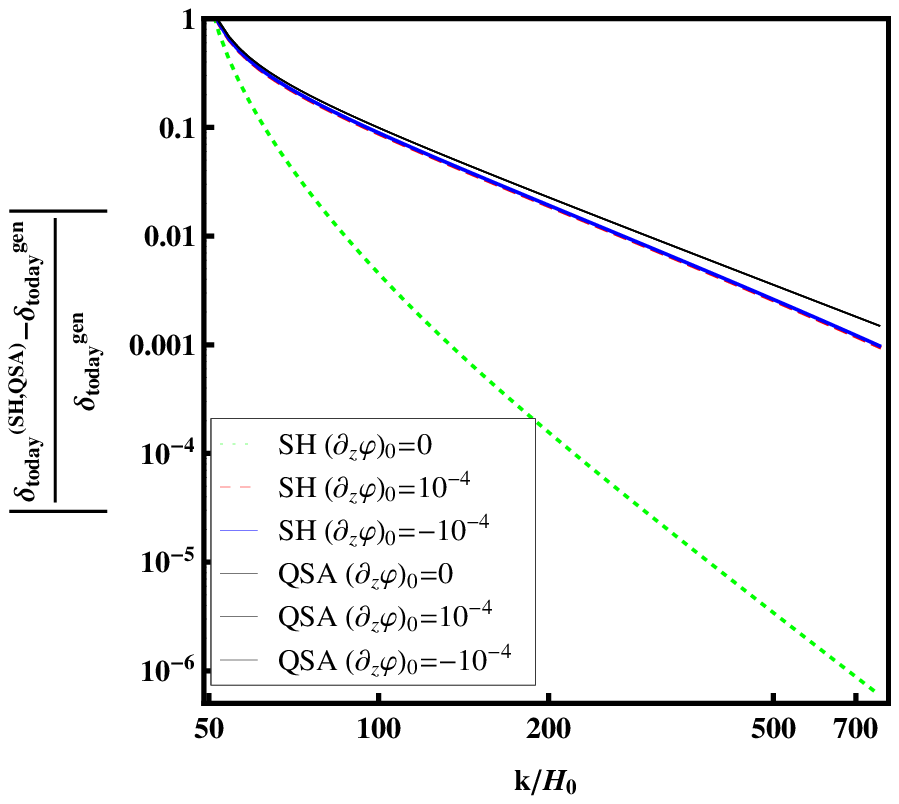}
	\caption{\footnotesize{Model $\alpha=10^{-3}$:
	Left panel: Value of $\delta_k$ today for $\partial_z\varphi = (-10^{-4},0,10^{-4})$, for a wide range of modes described by the general equation and $\Lambda$CDM.
	Centre panel: Relative difference of the value of $\delta_k$ today with respect to the value predicted by the $\Lambda$CDM model. Depending on the $\partial_z\varphi$ initial condition oscillations in such difference can be observed.
	Right panel: Relative error when using sub-Hubble or quasistatic approximations instead of the general equation. For the sub-Hubble approximation, the deviation with respect to
	$\Lambda$CDM strongly depends on the $\varphi'$ initial conditions. $\delta(z=1100)=10^{-3}$ and $\delta'(z=1100)=0$ have been used as initial conditions in all cases.}}
	\label{C1-1000deltatoday}		
\end{figure*}

Finally in Figure \ref{C1-1000deltatoday} (left panel) we plot the density contrast evaluated today as a function of $k$ for the initial conditions under consideration.  In the centre panel we then depict  the relative difference between the today values as predicted by JFBD theories \eqref{ecdeltagen} and by $\Lambda$CDM \eqref{ecdeltasRG}. As can be seen in the centre panel the value predicted by JFBD theories oscillate around the $\Lambda$CDM values, and the relative difference depends on the initial condition for $\varphi'$. The difference between $\Lambda$CDM and the case with null initial condition for $\varphi'$ is of order $10^{-5}$. We can deduce from these facts that in this case all the difference with $\Lambda$CDM is ruled by the scalar field $\varphi$ while the coupling $\alpha$ hardly produces any difference. Accordingly,  for this coupling any possible discrepancy with respect to the $\Lambda$CDM fittings in the processed matter power spectra is expected to be negligible.
In Figure \ref{C1-1000deltatoday} right panel we depict the relative error of either the quasistatic approximation \eqref{ecdeltaQSA} or sub-Hubble \eqref{ecdeltaSH} approximations
with respect to the general solution given by  \eqref{ecdeltagen}. It is shown how the sub-Hubble approximation is always a better approximation than the quasistatic approximation. The $k$-modes when described by the sub-Hubble approximation have a strong dependence on the $\varphi'$ initial conditions whereas the quasistatic evolution
hardly depends on $\varphi'$.


\subsection{Case $\alpha=10^{-1}$}

For this case, we have also checked that the scalar contribution to the background  stress-energy tensor is always negligible (lower than $1\%$ of the standard $\Lambda$CDM at any time in the evolution). Therefore, background approximations
are also justified in this case.

\begin{figure*} [htbp] 
	\centering
		\includegraphics[width=0.309\textwidth]{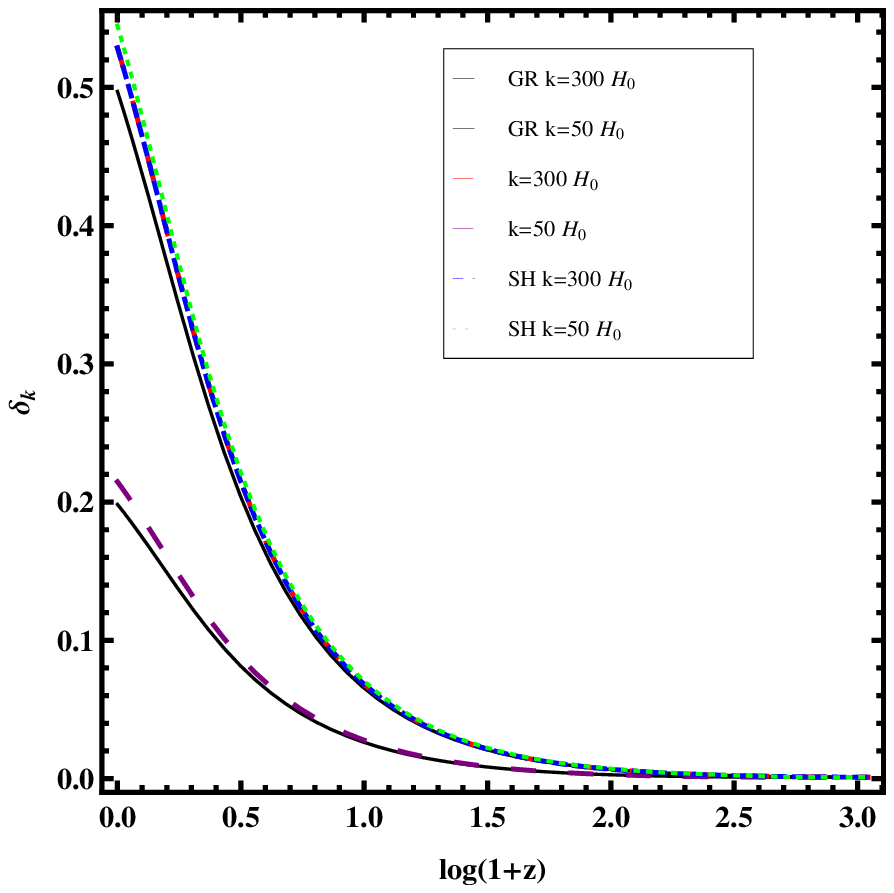}
		\includegraphics[width=0.329\textwidth]{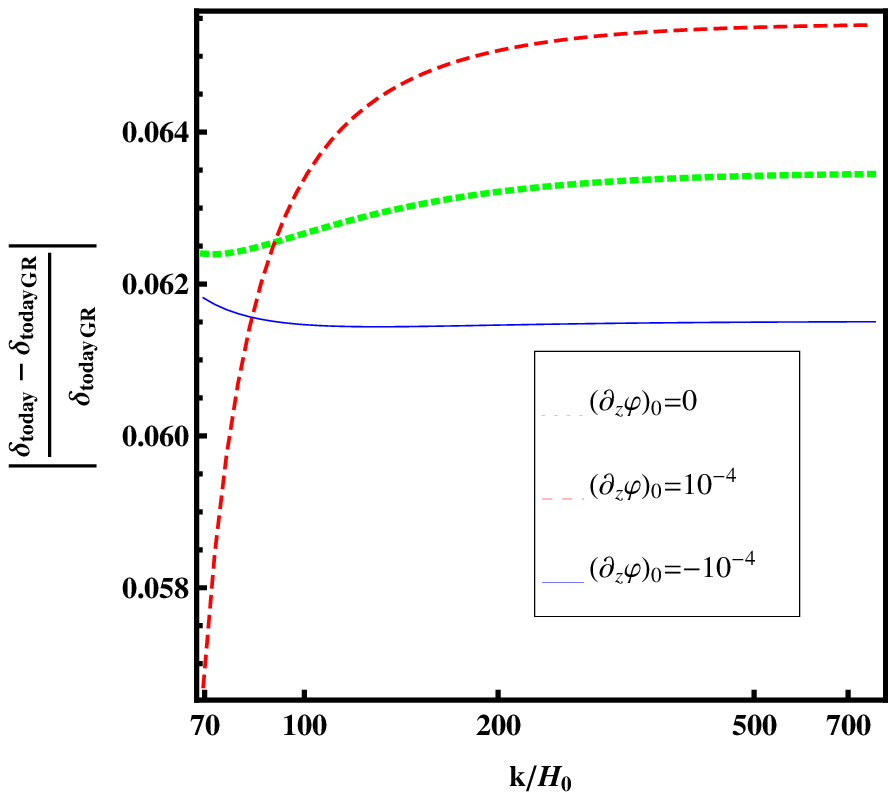}
		\includegraphics[width=0.329\textwidth]{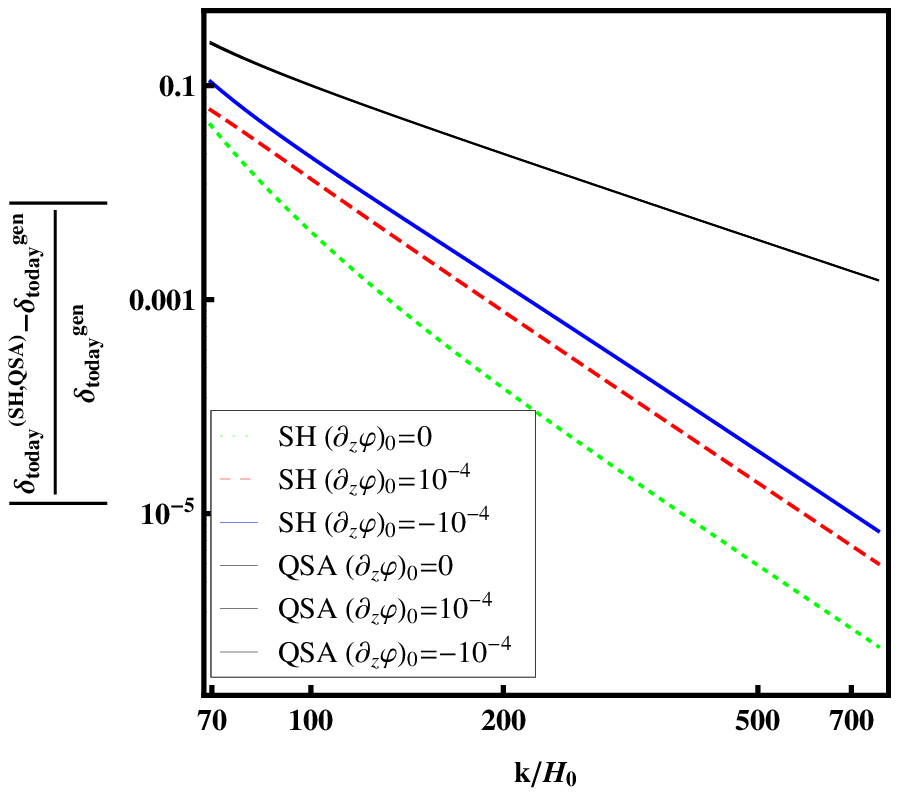}
 \caption{\footnotesize{Model $\alpha=10^{-1}$:
	 Left panel: $\delta_k$ for $\partial_z\varphi_0=0$, for the modes $k\;=\;50\mathcal{H}_0,\,300\mathcal{H}_0$ as described by the general equation \eqref{ecdeltagen}, the sub-Hubble approximation \eqref{ecdeltaSH} and $\Lambda$CDM \eqref{ecdeltasRG} in the redshift range from 1100 to 0. We can see that $\Lambda$CDM and the general equation predict different evolutions and that the sub-Hubble approximation represents a good approximation of the general equation for the highest mode ($k\,=\,300\mathcal{H}_0$).
	Centre panel: relative difference on the value of $\delta_k$ today, predicted by the models with $\partial_z\varphi = (-10^{-4},0,10^{-4})$, with respect to the value predicted by the $\Lambda$CDM model. The deviation is almost constant for every mode $k$ with a slight dependence on the initial condition for $\partial_z\varphi$.
	Right panel: Relative error when using quasistatic or sub-Hubble approximation instead of the general equation. For the sub-Hubble approximation, the relative deviation depends
	on the $\varphi'$ initial conditions. $\delta(z=1100)=10^{-3}$ and $\delta'(z=1100)=0$ have been used as initial conditions in all cases.}}
 \label{C1-10}
\end{figure*}

We present in Figure \ref{C1-10} a brief summary of the obtained results for this case.
We plotted there the density contrast evolution as provided by the general equation, sub-Hubble approximation and $\Lambda$CDM for two $k-$modes ($k\;=\;50\mathcal{H}_{0},\,300\mathcal{H}_0$). Thus we can see how for higher values of  the coupling $\alpha$, the JFBD density contrast evolution
becomes more different from the $\Lambda$CDM evolution.
The centre panel in Figure \ref{C1-10} shows the relative difference of $\delta$ evaluated today from the general equation when compared with $\Lambda$CDM values. This
difference turns out to be of the order $6\%$ and consequently bigger than in the previous case (smaller coupling, $\alpha=10^{-3}$). Therefore, the growth of perturbations in this model is mainly ruled by the coupling $\alpha$ with a slight dependence on the initial condition for $\varphi'$.
Finally,  we can also observe in this figure (right panel) that the sub-Hubble approximation provides a better description of the $\delta$ evolution than the quasistatic approximation. In fact, for this value
of $\alpha$, the sub-Hubble approximation is able to describe accurately a wider range of $k-$modes than with a smaller coupling regardless of the initial conditions for $\varphi'$.
%


\section{Conclusions}
\label{V}

In this work we have studied the evolution of matter density perturbations in Jordan-Fierz-Brans-Dicke theories in the Einstein Frame. We have presented for the first time in the available literature a completely general procedure to obtain the exact differential equation for the evolution of density contrast departing from
an Einstein-Hilbert gravitational action with cosmological constant and supplemented by a
scalar field with kinematical term and a coupling between standard matter and the scalar field.

Firstly, we have derived the modified Einstein equations in these kinds of theories both for the background and for the first order perturbed equations. Then, after algebraically combining the involved equations and without performing any intermediate simplification, we have shown that the fully general density perturbations equation is a fourth order differential equation. This is a key result of this investigation which is in agreement with the usual results for other modified gravity theories, such as $f(R)$ theories, for which the density contrast evolution equation is also fourth order. 
In the absence of the scalar field the general equation as predicted by General Relativity is naturally recovered.

For these kinds of theories, we have also studied the so-called sub-Hubble approximation equation which is valid for sub-Hubble modes. This equation turns out to be fourth order and $k$-mode dependent,  in contrast with its counterpart in General Relativity which is second-order and $k$-mode independent.
We recover as well, but this time from the completely valid equation, the quasistatic approximation presented
in other works in the literature both in the Einstein Frame and the Jordan Frame.


In order to illustrate our results, we have chosen two couplings, $\alpha=10^{-3},\, 10^{-1}$, between the scalar field and standard matter. The former is an allowed value for $\alpha$ whereas the latter was chosen to illustrate the parameter dependence. We have then studied the evolution of the perturbations for different initial values for the scalar field and compared the sub-Hubble evolution and the full equation in order to determine the validity of such approximation depending on the $k$-mode under consideration. The sub-Hubble approximation was proved to be valid for different scales and generally covers a wider range than that described by the quasistatic approximation. 

For each coupling and three different initial conditions for the scalar field first derivative, we also compared the general evolution with respect to that predicted by the Concordance $\Lambda$CDM model. We concluded that for allowed parameters $\alpha$ the relative difference today is always lower than 5\%. We have also shown that as the coupling $\alpha$ becomes bigger the difference with $\Lambda$CDM increases. 
Furthermore we could observe that when the coupling $\alpha$ is small, the difference with $\Lambda$CDM  is mainly driven by the scalar field.

The implemented method for solving the perturbed equations is general and could be used to study the perturbed equations for other scalar-tensor theories with non-constant coupling and/or with scalar field potential. Further work in this direction is in progress. Thus, it should be possible to use this general approach to constrain which scalar-tensor theories continue to be consistent with matter power spectra current data even if at the level of the FLRW cosmological background they remain indistinguishable from $\Lambda$CDM model.


\

{\bf Acknowledgements:}
JARC and AdlCD acknowledge financial support from MINECO (Spain) projects numbers FIS2011-23000, FPA2011-27853-C02-01 and Consolider-Ingenio MULTIDARK CSD2009-00064.
AdlCD also acknowledges financial support from Marie Curie - Beatriu de Pin\'os contract BP-B00195, Generalitat de Catalunya and ACGC, University of Cape Town.
LOG is indebted to the Theoretical Physics Department, Complutense University of Madrid for the use of technical resources.

\newpage
\appendix

\begin{widetext}
\section{Coefficients of the differential equation of the evolution of density perturbations}
\label{APPENDIX A}

In this Appendix we show the coefficients $\mathcal{C}_{0,1,2,3,4}$ 
of equations \eqref{ecdeltagen} and \eqref{ecdeltaSH} as a sum of powers in the $\epsilon$ coefficient (where $\epsilon\equiv \mathcal{H}/k$), i.e., $\mathcal{C}_i = \sum_{n_{min}}^{4}{\left(\epsilon/\mathcal{H}\right)^{2n}\mathcal{C}_{i,2n}}$,
where $n_{min} = 0$ for $\mathcal{C}_{0,1,2}$ 
and $n_{min} = 1$ for $\mathcal{C}_{3,4}$.

\begin{itemize}
\item Coefficients for $\delta$ term:

\begin{eqnarray}
\mathcal{C}_{0,0}=-4\alpha ^2 \left(\alpha ^2+1\right) \tilde{\rho},
\end{eqnarray}

\begin{eqnarray}
\mathcal{C}_{0,2}=-\left(\alpha ^2+1\right) \tilde{\rho}\left\{\left[7 \mathcal{H}^2+2 \left(5 \alpha ^2-7\right) \tilde{\rho}\right] \alpha ^2+2 \left(15 \alpha ^2+34\right) \mathcal{H} \varphi'\alpha +\left(-7 \alpha ^4-50 \alpha ^2+16\right) \varphi'^2\right\},
\end{eqnarray}

\begin{eqnarray}
\mathcal{C}_{0,4}&=&\tilde{\rho}\left\{\left(2 \alpha ^8-30 \alpha ^6-53 \alpha ^4+55 \alpha ^2+276\right) \varphi'^4+2 \alpha  \left(6 \alpha ^6+24 \alpha^4+23 \alpha ^2+205\right) \mathcal{H} \varphi'^3\right.\nonumber\\
& &+\left.\left[2 \left(\alpha ^8-23 \alpha ^6-29 \alpha ^4+45 \alpha ^2+138\right) \tilde{\rho}-\left(37 \alpha ^6-153 \alpha ^4+495 \alpha ^2+237\right) \mathcal{H}^2\right]\varphi'^2\right.\nonumber\\
& &-\left.2 \alpha  \mathcal{H} \left[\left(6 \alpha ^4+39 \alpha ^2+57\right) \mathcal{H}^2+\left(6 \alpha ^6-109 \alpha ^4+160 \alpha ^2-285\right) \tilde{\rho}\right] \varphi'\right.\nonumber\\
& &+\left.\alpha ^2 \left[-\left(\alpha ^2+57\right)\mathcal{H}^4-2 \left(7 \alpha ^4+24 \alpha ^2-39\right) \tilde{\rho} \mathcal{H}^2-4 \left(\alpha ^2-3\right) \left(\alpha ^4-\alpha ^2+14\right) \tilde{\rho}^2\right]\right\},
\end{eqnarray}

\begin{eqnarray}
\mathcal{C}_{0,6}&=&\tilde{\rho}\left\{\left(-10 \alpha ^8+101 \alpha^6-66 \alpha ^4+245 \alpha ^2-1224\right) \varphi'^6-2 \alpha  \left(12 \alpha ^6-158 \alpha ^4+185 \alpha ^2+399\right) \mathcal{H} \varphi'^5\right.\nonumber\\
& &+\left.\left[\left(-59 \alpha ^6+51 \alpha ^4-2795 \alpha ^2+4437\right)\mathcal{H}^2-2 \left(19 \alpha ^8-201 \alpha ^6+269 \alpha ^4-261 \alpha ^2+1332\right) \tilde{\rho}\right] \varphi'^4\right.\nonumber\\
& &+\left.2 \alpha  \mathcal{H} \left[2 \left(29 \alpha ^4-427 \alpha ^2+864\right)\mathcal{H}^2+\left(6 \alpha ^6+135\alpha ^4-772 \alpha ^2+483\right) \tilde{\rho}\right] \varphi'^3\right.\nonumber\\
& &+\left.\left[\left(441 \alpha ^4+1510 \alpha ^2-765\right) \mathcal{H}^4+4 \left(9 \alpha ^6-519 \alpha ^4+731 \alpha ^2+657\right) \tilde{\rho} \mathcal{H}^2\right.\right.\nonumber\\
& &-\left.\left.2 \left(22\alpha ^8-197 \alpha ^6+382 \alpha ^4-263 \alpha ^2+720\right) \tilde{\rho}^2\right] \varphi'^2\right.\nonumber\\
& &+\left.2 \alpha  \mathcal{H} \left[-\left(\alpha ^2+357\right) \mathcal{H}^4+\left(65 \alpha ^4+194 \alpha ^2+537\right) \tilde{\rho}\mathcal{H}^2+\left(39 \alpha ^6-455 \alpha ^4+977 \alpha ^2-225\right) \tilde{\rho}^2\right] \varphi'\right.\nonumber\\
& &+\left.2 \alpha ^2 \left[3 \mathcal{H}^6+6 \left(4 \alpha ^2-1\right) \tilde{\rho} \mathcal{H}^4+\left(-35 \alpha ^4+92 \alpha^2+207\right) \tilde{\rho}^2 \mathcal{H}^2-8 \left(\alpha ^6-15 \alpha ^2+18\right) \tilde{\rho}^3\right]\right\},
\end{eqnarray}

\begin{eqnarray}
\mathcal{C}_{0,8}&=&2 \tilde{\rho} \left(3 \mathcal{H}^2-\varphi'^2-2 \tilde{\rho}\right) \left\{\left(-4 \alpha ^8+60 \alpha ^6-165\alpha ^4+354 \alpha ^2-567\right) \varphi'^6+\alpha  \left(-14 \alpha ^6+167 \alpha ^4-724 \alpha ^2+945\right) \mathcal{H} \varphi'^5\right.\nonumber\\
& &+\left.\left[\left(-30 \alpha ^6+133 \alpha ^4-1818 \alpha ^2+3159\right) \mathcal{H}^2-2\left(8 \alpha ^8-124 \alpha ^6+411 \alpha ^4-624 \alpha ^2+567\right) \tilde{\rho}\right] \varphi'^4\right.\nonumber\\
& &+\left.\alpha  \mathcal{H} \left[\left(89 \alpha ^4-316 \alpha ^2+1215\right) \mathcal{H}^2-4 \left(7 \alpha ^6-86 \alpha ^4+529\alpha ^2-810\right) \tilde{\rho}\right] \varphi'^3\right.\nonumber\\
& &+\left.\left[2 \left(8 \alpha ^4+141 \alpha ^2-162\right) \mathcal{H}^4+2 \left(-9 \alpha ^6-154 \alpha ^4+225 \alpha ^2+486\right) \tilde{\rho} \mathcal{H}^2\right.\right.\nonumber\\
& &+\left.\left.\left(-22 \alpha ^8+315\alpha ^6-1155 \alpha ^4+1461 \alpha ^2-567\right) \tilde{\rho}^2\right] \varphi'^2\right.\nonumber\\
& &+\left.3 \alpha  \mathcal{H} \left[2 \alpha ^2 \mathcal{H}^4+8 \left(3 \alpha ^4+8 \alpha ^2+9\right) \tilde{\rho} \mathcal{H}^2+\left(-5 \alpha ^6+5 \alpha^4-95 \alpha ^2+279\right) \tilde{\rho}^2\right] \varphi'\right.\nonumber\\
& &+\left.12 \alpha ^2 \tilde{\rho} \left[-\alpha ^2 \mathcal{H}^4+\left(-\alpha ^4+\alpha ^2+18\right) \tilde{\rho} \mathcal{H}^2-\left(\alpha ^2-3\right)^3\tilde{\rho}^2\right]\right\}.
\end{eqnarray}

\item Coefficients for $\delta'$ term:

\begin{eqnarray}
\mathcal{C}_{1,0}=4\alpha ^2 \left(\mathcal{H}+\alpha\varphi'\right),
\end{eqnarray}

\begin{eqnarray}
\mathcal{C}_{1,2}&=&\left\{\alpha\left(\alpha ^4-38 \alpha ^2+16\right) \varphi'^3+\left(15 \alpha ^4+22 \alpha ^2+16\right) \mathcal{H} \varphi'^2+\alpha  \left[2 \left(7\alpha^2+11\right) \tilde{\rho} \alpha ^2+\left(9 \alpha ^2+68\right) \mathcal{H}^2\right] \varphi'\right.\nonumber\\
& &+\left.\alpha ^2 \mathcal{H} \left[19 \mathcal{H}^2+6 \left(\alpha ^2-3\right) \tilde{\rho}\right]\right\},
\end{eqnarray}

\begin{eqnarray}
\mathcal{C}_{1,4}&=&\left\{\alpha  \left(-11 \alpha ^4+181\alpha ^2-264\right) \varphi'^5-\left(79 \alpha ^4+57 \alpha ^2+324\right) \mathcal{H} \varphi'^4\right.\nonumber\\
& &+\left.\alpha  \left[\left(2 \alpha ^4+305 \alpha ^2-277\right) \mathcal{H}^2+\left(3 \alpha ^6-103 \alpha ^4+210 \alpha^2-272\right) \tilde{\rho}\right] \varphi'^3\right.\nonumber\\
& &+\left.\mathcal{H} \left[\left(2 \alpha ^4-233 \alpha ^2+413\right) \mathcal{H}^2+\left(19 \alpha ^6+175 \alpha ^4-148 \alpha ^2-420\right) \tilde{\rho}\right] \varphi'^2\right.\nonumber\\
& &+\left.\alpha \left[2 \left(17 \alpha ^2+99\right) \mathcal{H}^4+\left(-11 \alpha ^4+243 \alpha ^2-446\right) \tilde{\rho} \mathcal{H}^2+2 \left(3 \alpha ^6-32 \alpha ^4+41 \alpha ^2-68\right) \tilde{\rho}^2\right] \varphi'\right.\nonumber\\
& &+\left.\alpha ^2 \mathcal{H}\left[-2 \mathcal{H}^4+7 \left(3 \alpha ^2-1\right) \tilde{\rho} \mathcal{H}^2+2 \left(\alpha ^4+96 \alpha ^2-145\right) \tilde{\rho}^2\right]\right\},
\end{eqnarray}

\begin{eqnarray}
\mathcal{C}_{1,6}&=&\left\{2 \alpha  \left(19 \alpha ^4-232 \alpha ^2+481\right) \varphi'^7+2\left(43 \alpha ^4-468 \alpha ^2+945\right) \mathcal{H} \varphi'^6\right.\nonumber\\
& &+\left.\alpha  \left[\left(-19 \alpha ^6+258 \alpha ^4-1591 \alpha ^2+2542\right) \tilde{\rho}-2 \left(27 \alpha ^4-424 \alpha ^2+741\right) \mathcal{H}^2\right]\varphi'^5\right.\nonumber\\
& &+\left.\mathcal{H} \left[\left(-294 \alpha ^4+2696 \alpha ^2-4722\right) \mathcal{H}^2+\left(-23 \alpha ^6+724 \alpha ^4-2681 \alpha ^2+3150\right) \tilde{\rho}\right] \varphi'^4\right.\nonumber\\
& &-\left.\alpha  \left[24 \left(\alpha^2+36\right) \mathcal{H}^4+\left(255 \alpha ^4-3662 \alpha ^2+6567\right) \tilde{\rho} \mathcal{H}^2+4 \left(11 \alpha ^6-120 \alpha ^4+549 \alpha ^2-605\right) \tilde{\rho}^2\right] \varphi'^3\right.\nonumber\\
& &+\left.\mathcal{H} \left[72 \left(3-5 \alpha^2\right) \mathcal{H}^4+\left(-505 \alpha ^4+758 \alpha ^2-3525\right) \tilde{\rho} \mathcal{H}^2+4 \left(24 \alpha ^6+375 \alpha ^4-982 \alpha ^2+558\right) \tilde{\rho}^2\right] \varphi'^2\right.\nonumber\\
& &+\left.2 \alpha  \tilde{\rho} \left[\left(116 \alpha ^2-243\right) \mathcal{H}^4-2 \left(116 \alpha ^4-221 \alpha ^2+558\right) \tilde{\rho} \mathcal{H}^2+2 \left(89 \alpha ^4-380 \alpha ^2+291\right) \tilde{\rho}^2\right] \varphi'\right.\nonumber\\
& &+\left.2 \alpha ^2 \mathcal{H} \tilde{\rho} \left[3 \mathcal{H}^4+2 \left(68 \alpha ^2-123\right) \tilde{\rho} \mathcal{H}^2-2 \left(\alpha ^4-20 \alpha ^2+3\right) \tilde{\rho}^2\right]\right\},
\end{eqnarray}

\begin{eqnarray}
\mathcal{C}_{1,8}&=&+2 \left\{-4 \alpha  \left(5 \alpha ^4-69 \alpha ^2+120\right) \varphi'^9-4 \left(17 \alpha ^4-115 \alpha ^2+270\right) \mathcal{H} \varphi'^8\right.\nonumber\\
& &+\left.\alpha  \left[8 \left(9 \alpha ^4-134 \alpha ^2+156\right) \mathcal{H}^2+\left(14 \alpha ^6-207 \alpha ^4+1278 \alpha ^2-1755\right)\tilde{\rho}\right] \varphi'^7\right.\nonumber\\
& &+\left.\mathcal{H} \left[8 \left(36 \alpha ^4-194 \alpha ^2+585\right) \mathcal{H}^2+\left(13 \alpha ^6-787 \alpha ^4+1889 \alpha ^2-2727\right) \tilde{\rho}\right] \varphi'^6\right.\nonumber\\
& &\left.+\alpha\left[12\left(-3 \alpha ^4+63 \alpha ^2+56\right) \mathcal{H}^4+2 \left(-3 \alpha ^6+253 \alpha ^4-1671 \alpha ^2+2724\right) \tilde{\rho} \mathcal{H}^2\right.\right.\nonumber\\
& &\left.\left.+\left(60 \alpha ^6-611 \alpha ^4+2325 \alpha ^2-2394\right) \tilde{\rho}^2\right]\varphi'^5\right.\nonumber\\
& &-\left.\mathcal{H} \left[36 \left(7 \alpha ^4-15 \alpha ^2+120\right) \mathcal{H}^4+\left(75 \alpha ^6-1595 \alpha ^4+1357 \alpha ^2-4365\right) \tilde{\rho} \mathcal{H}^2\right.\right.\nonumber\\
& &+\left.\left.\left(12 \alpha ^8-11 \alpha ^6+1451 \alpha ^4-3082 \alpha^2+2538\right) \tilde{\rho}^2\right] \varphi'^4-\alpha  \left[72 \left(\alpha ^2+4\right) \mathcal{H}^6+\left(555 \alpha ^4-2936 \alpha ^2+8607\right) \tilde{\rho} \mathcal{H}^4\right.\right.\nonumber\\
& &+\left.\left.\left(36 \alpha ^6-1221 \alpha ^4+2422 \alpha^2-3741\right) \tilde{\rho}^2 \mathcal{H}^2+\left(-82 \alpha ^6+709 \alpha ^4-2046 \alpha ^2+1659\right) \tilde{\rho}^3\right] \varphi'^3\right.\nonumber\\
& &-\left.3 \mathcal{H} \left[24 \alpha ^2 \mathcal{H}^6+2 \left(19 \alpha ^4+217 \alpha ^2-198\right) \tilde{\rho} \mathcal{H}^4+\left(61 \alpha ^6-934 \alpha ^4+2385 \alpha ^2-192\right) \tilde{\rho}^2 \mathcal{H}^2\right.\right.\nonumber\\
& &+\left.\left.\left(10 \alpha ^8+137 \alpha ^4-404 \alpha ^2+297\right) \tilde{\rho}^3\right] \varphi'^2-3 \alpha  \tilde{\rho}\left[\left(8 \alpha ^2-30\right) \mathcal{H}^6+\left(132 \alpha ^4+209 \alpha ^2-321\right) \tilde{\rho} \mathcal{H}^4\right.\right.\nonumber\\
& &+\left.\left.\left(\alpha ^6-446 \alpha ^4+943 \alpha ^2-162\right) \tilde{\rho}^2 \mathcal{H}^2-4 \left(\alpha ^2-3\right)^2 \left(3\alpha ^2-5\right) \tilde{\rho}^3\right] \varphi'\right.\nonumber\\
& &+\left.3 \alpha ^2 \mathcal{H} \tilde{\rho}^2 \left[\left(57 \alpha ^2-45\right) \mathcal{H}^4+\left(-31 \alpha ^4-132 \alpha ^2+171\right) \tilde{\rho} \mathcal{H}^2-4 \left(\alpha^2-3\right)^2 \left(\alpha ^2+1\right) \tilde{\rho}^2\right]\right\}.
\end{eqnarray}

\item Coefficients for $\delta''$ term:

\begin{eqnarray}
\mathcal{C}_{2,0}=4 \alpha ^2,
\end{eqnarray}

\begin{eqnarray}
\mathcal{C}_{2,2}=\left\{\left[19 \mathcal{H}^2+2 \left(\alpha ^2-3\right) \tilde{\rho}\right] \alpha ^2+2 \left(3 \alpha ^2+34\right) \mathcal{H} \varphi' \alpha +\left(\alpha ^4-54 \alpha ^2+16\right) \varphi'^2\right\},
\end{eqnarray}

\begin{eqnarray}
\mathcal{C}_{2,4}&=&\left\{\left(-9 \alpha ^4+327 \alpha ^2-292\right) \varphi'^4+2 \alpha  \left(\alpha ^4-61 \alpha ^2-203\right) \mathcal{H} \varphi'^3\right.\nonumber\\
& &+\left.\left[\left(14 \alpha ^4-139 \alpha ^2+413\right) \mathcal{H}^2-4 \left(20 \alpha^4-59 \alpha ^2+61\right) \tilde{\rho}\right] \varphi'^2+2 \alpha  \mathcal{H} \left[\left(20 \alpha ^2+99\right) \mathcal{H}^2+\left(2 \alpha ^4+57 \alpha ^2-85\right) \tilde{\rho}\right] \varphi'\right.\nonumber\\
& &+\left.2 \alpha^2\left[-\mathcal{H}^4+\left(11 \alpha ^2-9\right) \tilde{\rho} \mathcal{H}^2+6 \left(\alpha ^2-3\right) \tilde{\rho}^2\right]\right\},
\end{eqnarray}

\begin{eqnarray}
\mathcal{C}_{2,6}&=&2\left\{\left(33 \alpha ^4-341 \alpha ^2+690\right) \varphi'^6+\alpha  \left(-15 \alpha^4+125 \alpha ^2-252\right) \mathcal{H} \varphi'^5\right.\nonumber\\
& &+\left.\left[\left(-90 \alpha ^4+1141 \alpha ^2-2361\right) \mathcal{H}^2+\left(-7 \alpha ^6+177 \alpha ^4-744 \alpha ^2+1182\right) \tilde{\rho}\right] \varphi'^4\right.\nonumber\\
& &+\left.\alpha  \mathcal{H}\left[\left(39 \alpha ^2-81\right) \mathcal{H}^2+\left(-55 \alpha ^4+378 \alpha ^2-945\right) \tilde{\rho}\right] \varphi'^3\right.\nonumber\\
& &+\left.\left[36 \left(3-5 \alpha ^2\right) \mathcal{H}^4+\left(-128 \alpha ^4+705 \alpha ^2-1245\right)\tilde{\rho} \mathcal{H}^2+\left(-20 \alpha ^6+139 \alpha ^4-409 \alpha ^2+492\right) \tilde{\rho}^2\right] \varphi'^2\right.\nonumber\\
& &+\left.\alpha  \mathcal{H} \tilde{\rho} \left[\left(220 \alpha ^2-438\right) \mathcal{H}^2+\left(-29 \alpha ^4+172 \alpha  ^2-303\right) \tilde{\rho}\right] \varphi'+3 \alpha ^2 \mathcal{H}^2 \tilde{\rho} \left[3 \left(13 \alpha ^2-31\right) \tilde{\rho}-28 \mathcal{H}^2\right]\right\},
\end{eqnarray}

\begin{eqnarray}
\mathcal{C}_{2,8}&=&2 \left\{\left(-25 \alpha ^4+375 \alpha ^2-756\right)  \varphi'^8+\alpha  \left(\alpha ^2-3\right) \left(8 \alpha ^2-105\right) \mathcal{H} \varphi'^7\right.\nonumber\\
& &+\left.\left[6 \left(23 \alpha ^4-343 \alpha ^2+606\right) \mathcal{H}^2+\left(3 \alpha ^6-230 \alpha ^4+1284 \alpha ^2-1890\right)  \tilde{\rho}\right] \varphi'^6\right.\nonumber\\
& &+\left.\alpha  \mathcal{H} \left[\left(-24 \alpha ^4+430 \alpha ^2-690\right) \mathcal{H}^2+\left(6 \alpha ^6+49 \alpha ^4-879 \alpha ^2+1872\right) \tilde{\rho}\right] \varphi'^5\right.\nonumber\\
& &+\left.\left[-3 \left(63\alpha ^4-949 \alpha ^2+1368\right) \mathcal{H}^4+3 \left(-5 \alpha ^6+242 \alpha ^4-1056 \alpha ^2+1236\right) \tilde{\rho} \mathcal{H}^2\right.\right.\nonumber\\
& &+\left.\left.\left(18 \alpha ^6-427 \alpha ^4+1467 \alpha ^2-1512\right) \tilde{\rho}^2\right]\varphi'^4+3 \alpha  \mathcal{H} \left[-\left(43 \alpha ^2+255\right) \mathcal{H}^4+\left(-99 \alpha ^4+832 \alpha ^2-1337\right) \tilde{\rho} \mathcal{H}^2\right.\right.\nonumber\\
& &+\left.\left.\left(2 \alpha ^6+57 \alpha ^4-258 \alpha ^2+375\right) \tilde{\rho}^2\right]  \varphi'^3-3 \left[48 \alpha ^2 \mathcal{H}^6+\left(58 \alpha ^4-185 \alpha ^2+288\right) \tilde{\rho} \mathcal{H}^4\right.\right.\nonumber\\
& &+\left.\left.2 \left(14 \alpha ^6-161 \alpha ^4+360 \alpha ^2-201\right) \tilde{\rho}^2 \mathcal{H}^2-2 \left(\alpha^2-3\right)^2\left(4 \alpha ^2-7\right) \tilde{\rho}^3\right] \varphi'^2\right.\nonumber\\
& &-\left.3 \alpha  \mathcal{H} \tilde{\rho} \left[3 \left(7 \alpha ^2+85\right) \mathcal{H}^4+\left(85 \alpha ^4-192 \alpha ^2+219\right) \tilde{\rho} \mathcal{H}^2+4 \left(\alpha^2-3\right)^2 \left(\alpha ^2-2\right) \tilde{\rho}^2\right] \varphi'\right.\nonumber\\
& &+\left.9 \alpha ^2 \mathcal{H}^2 \tilde{\rho} \left[\mathcal{H}^4+\left(15 \alpha ^2-49\right) \tilde{\rho} \mathcal{H}^2-2 \left(\alpha ^2-3\right)^2\tilde{\rho}^2\right]\right\}.
\end{eqnarray}

\item Coefficients for $\delta'''$ term:

\begin{eqnarray}
\mathcal{C}_{3,2}=4\alpha^2\left(\alpha  \varphi'+3 \mathcal{H}\right),
\end{eqnarray}

\begin{eqnarray}
\mathcal{C}_{3,4}&=&\left\{\alpha ^2 \mathcal{H} \left[2 \left(7 \alpha ^2+5\right) \tilde{\rho}-\mathcal{H}^2\right]+\alpha  \varphi' \left[\left(23 \alpha ^2+144\right) \mathcal{H}^2+2\left(\alpha ^4-23 \alpha ^2+34\right) \tilde{\rho}\right]\right.\nonumber\\
& &+\left.\left(13 \alpha ^4-148 \alpha ^2+112\right) \mathcal{H} \varphi'^2+\alpha  \left(\alpha ^4-20 \alpha ^2+52\right) \varphi'^3\right\},
\end{eqnarray}

\begin{eqnarray}
\mathcal{C}_{3,6}&=&\left\{2 \alpha ^2\mathcal{H} \tilde{\rho} \left[2 \left(29 \alpha ^2-75\right) \tilde{\rho}-45 \mathcal{H}^2\right]+\alpha  \varphi'^3 \left[7 \left(5 \alpha ^2-19\right) \mathcal{H}^2-2 \left(10 \alpha ^4-138 \alpha ^2+253\right)\tilde{\rho}\right]\right.\nonumber\\
& &+\left.\mathcal{H} \varphi'^2 \left[5 \left(69-35 \alpha ^2\right) \mathcal{H}^2-2 \left(28 \alpha ^4-80 \alpha ^2+363\right) \tilde{\rho}\right]\right.\nonumber\\
& &-\left.2 \alpha  \varphi' \left[3 \left(42-37 \alpha ^2\right) \mathcal{H}^2\tilde{\rho}+2 \left(8 \alpha ^4-43 \alpha ^2+69\right) \tilde{\rho}^2+30 \mathcal{H}^4\right]\right.\nonumber\\
& &+\left.\left(-47 \alpha ^4+511 \alpha ^2-1050\right) \mathcal{H} \varphi'^4+\alpha  \left(-5 \alpha ^4+61 \alpha ^2-122\right)\varphi'^5\right\},
\end{eqnarray}

\begin{eqnarray}
\mathcal{C}_{3,8}&=&2\left\{3 \alpha ^2 \mathcal{H} \tilde{\rho} \left[3 \left(9 \alpha ^2-31\right) \mathcal{H}^2 \tilde{\rho}-4 \left(\alpha ^2-3\right)^2 \tilde{\rho}^2+3 \mathcal{H}^4\right]\right.\nonumber\\
& &+\left.\alpha  \varphi'^5 \left[3 \left(2\alpha ^4-49 \alpha ^2+86\right) \tilde{\rho}-2 \left(3 \alpha ^4-56 \alpha ^2+117\right) \mathcal{H}^2\right]\right.\nonumber\\
& &+\left.\mathcal{H} \varphi'^4 \left[\left(129 \alpha ^4-568 \alpha ^2+1170\right) \tilde{\rho}-3 \left(21 \alpha ^4-280\alpha ^2+603\right) \mathcal{H}^2\right]\right.\nonumber\\
& &+\left.\alpha  \varphi'^3 \left[\left(351-93 \alpha ^2\right) \mathcal{H}^4+\left(-78 \alpha ^4+655 \alpha ^2-1593\right) \mathcal{H}^2 \tilde{\rho}+\left(14 \alpha ^4-195 \alpha ^2+327\right)\tilde{\rho}^2\right]\right.\nonumber\\
& &-\left.3 \alpha\tilde{\rho} \varphi' \left[\left(26 \alpha ^2+165\right) \mathcal{H}^4-4 \left(\alpha ^2-3\right)^2 \tilde{\rho}^2+\left(35 \alpha ^4-121 \alpha ^2+156\right) \mathcal{H}^2\tilde{\rho}\right]+\left(21 \alpha ^4-268 \alpha^2+639\right) \mathcal{H} \varphi'^6\right.\nonumber\\
& &+\left.\alpha  \left(2 \alpha ^4-27 \alpha ^2+39\right) \varphi'^7\right.\nonumber\\
& &-\left.3 \mathcal{H} \varphi'^2 \left[36 \left(\alpha ^2+3\right) \mathcal{H}^4+5\left(3 \alpha ^4-43 \alpha ^2+36\right) \mathcal{H}^2 \tilde{\rho}+\left(2 \alpha ^6-76 \alpha ^4+185 \alpha ^2-177\right) \tilde{\rho}^2\right]\right\}.
\end{eqnarray}

\item Coefficients for $\delta^{iv}$ term:

\begin{eqnarray}
\mathcal{C}_{4,2}=4 \alpha ^2,
\end{eqnarray}

\begin{eqnarray}
\mathcal{C}_{4,4}&=&+ \left\{6 \left(\alpha ^2+6\right) \alpha  \mathcal{H} \varphi'+\alpha ^2 \left[2 \left(\alpha^2-3\right) \tilde{\rho}-\mathcal{H}^2\right]+\left(\alpha ^4-26 \alpha ^2+16\right) \varphi'^2\right\},
\end{eqnarray}

\begin{eqnarray}
\mathcal{C}_{4,6}&=&+\left\{2\alpha\mathcal{H} \varphi' \left[2 \left(14 \alpha ^2-15\right) \tilde{\rho}-15 \mathcal{H}^2\right]+2 \alpha  \left(4 \alpha ^2-7\right) \mathcal{H} \varphi'^3+6 \alpha ^2 \tilde{\rho} \left[2 \left(\alpha ^2-3\right)\tilde{\rho}-5 \mathcal{H}^2\right]\right.\nonumber\\
& &+\left.\varphi'^2 \left[\left(69-29 \alpha ^2\right) \mathcal{H}^2-2 \left(5 \alpha ^4-20 \alpha ^2+54\right) \tilde{\rho}\right]+\left(-5 \alpha ^4+53 \alpha ^2-108\right)\varphi'^4\right\},
\end{eqnarray}

\begin{eqnarray}
\mathcal{C}_{4,8}&=&+2\left\{3 \alpha\left(\alpha ^2-8\right) \mathcal{H} \varphi'^5+9 \alpha ^2 \mathcal{H}^2 \tilde{\rho} \left[\left(3 \alpha ^2-5\right) \tilde{\rho}-\mathcal{H}^2\right]\right.\nonumber\\
& &+\left.\varphi'^4 \left[\left(-6 \alpha ^4+103 \alpha^2-153\right) \mathcal{H}^2+\left(11 \alpha ^4-72 \alpha ^2+126\right) \tilde{\rho}\right]-3 \alpha  \mathcal{H} \varphi'^3 \left[\left(3 \alpha ^2-23\right) \mathcal{H}^2+2 \left(\alpha ^4-5 \alpha ^2+16\right) \tilde{\rho}\right]\right.\nonumber\\
& &-\left.3\varphi'^2 \left[\left(13 \alpha ^2+36\right) \mathcal{H}^4+\left(\alpha ^4-35 \alpha ^2+6\right) \mathcal{H}^2 \tilde{\rho}+\left(-4 \alpha ^4+15 \alpha ^2-21\right) \tilde{\rho}^2\right]\right.\nonumber\\
& &-\left.3 \alpha  \mathcal{H} \varphi' \left[2\left(2 \alpha ^4-3 \alpha ^2+3\right) \tilde{\rho}^2+36 \mathcal{H}^2 \tilde{\rho}-3 \mathcal{H}^4\right]+\left(2 \alpha ^4-30 \alpha ^2+63\right) \varphi'^6\right\}.
\end{eqnarray}
\end{itemize}

\end{widetext}

\end{document}